\documentclass[epj]{svjour}
\usepackage[english]{babel}
\usepackage{graphicx}
\usepackage{bm}
\usepackage{amsmath}
\newcommand{\gras}[1]{\boldsymbol{#1}}
\newcommand{\same}[1]{\textbf{#1}}
\newcommand{\xh}{\hat{\mathbf{x}}}
\newcommand{\pounders}{\codes{POUNDERS}}

\newcommand{\codes}[1]{\textsf{#1}}
\begin{document}
\hyphenation{Baye-sian}
\hyphenation{back-ward}
\hyphenation{pa-ra-me-tri-za-tion}
\hyphenation{pa-ra-me-tri-za-tions}
\hyphenation{phe-no-me-no-lo-gi-cal}
\hyphenation{semi-phe-no-me-no-lo-gi-cal}
\hyphenation{Hohen-berg-Kohn}
\title{Uncertainty Quantification and Propagation in Nuclear Density 
Functional Theory}
\author{N. Schunck \inst{1} \and 
J.D. McDonnell \inst{1,2} \and 
D. Higdon \inst{3}
\and J. Sarich \inst{4} \and 
S.M. Wild \inst{4}
}
\institute{Nuclear and Chemical Science Division, Lawrence Livermore National 
Laboratory, Livermore, CA 94551, USA\and
Department of Physics and Astronomy, Francis Marion University,
Florence, SC 29501, USA\and
Los Alamos National Laboratory, Los Alamos, NM 87545, USA\and
Mathematics and Computer Science Division, Argonne National
Laboratory, Argonne, IL 60439, USA}
\date{Received: date / Revised version: date}
%
\abstract{
Nuclear density functional theory (DFT) is one of the main theoretical tools used 
to study the properties of heavy and superheavy elements, or to describe the 
structure of nuclei far from stability. While on-going efforts seek to 
better root nuclear DFT in the theory of nuclear forces [see Duguet {\it et al.}, 
this issue], energy functionals remain semi-phenomenological constructions that 
depend on a set of parameters adjusted to experimental data in finite nuclei. 
In this paper, we review recent efforts to quantify the related uncertainties, and 
propagate them to model predictions. In particular, we cover the topics of 
parameter estimation for inverse problems, statistical analysis of model 
uncertainties and Bayesian inference methods. Illustrative examples are taken 
from the literature.
} 
\maketitle

%
%
%
%
\section{Introduction}
\label{sec:introduction}

Applications of nuclear science in energy production or national security are 
based on nuclear data such as cross-sections, energy levels, and lifetimes. In 
many cases of interest, experimental measurements are not available, and guidance 
from theory is indispensable. In the valley of stability, one still can 
employ simple models heavily tuned to existing data. For example, the fission 
model implemented in the GEF code uses about 50 parameters such as fission 
barrier heights and level density parameter which are parameterized as a 
function of $Z$, $N$, or neutron incident energy. The code also uses databases 
of binding energies and shell corrections (in the ground state only). Based on 
these parameters and data banks, qualitative arguments, and a Monte Carlo 
sampling scheme, observables such as fission probabilities, fission fragment 
yields, and neutron multiplicities can be reproduced accurately in the actinide 
region (with a few exceptions) \cite{schmidt2014}. While such an empirical 
approach fulfills some of the needs of data evaluators, however, its predictive power 
beyond the region where the model is fitted is null. Indeed, such models do not 
contain any physics principle related to nucleons, their interaction, and the 
quantum nature of the atomic nucleus.

Therefore, even if data-driven empirical models will always be helpful in the 
short term, one must try to root data evaluation into more microscopic 
theories of nuclear structure and reactions in order to gain confidence in the 
reliability of evaluations. In heavy elements, density functional theory is 
currently the only candidate for such a microscopic approach to nuclear structure. 
In particular, recent advances in high-performance computing have enabled large-scale calculations of nuclear properties at the scale of the mass table 
\cite{erler2012,bogner2013}. Despite this progress, however, the accuracy and precision needed in data evaluations represent a formidable 
challenge for nuclear density functional theory (DFT). As an example, nuclear binding energies are computed 
within approximately 500 keV in state-of-the-art DFT calculations 
\cite{goriely2013,goriely2009}. Although this represent a relative error of 
0.05\% or less for nuclei with mass $A >100 $, it remains far from the sub-keV 
accuracy that is demanded in, for example, criticality studies. In order to make data 
evaluations based on microscopic inputs from DFT a viable alternative 
to simpler models, two challenges must be addressed in the next few years. 

First, DFT must be more firmly and 
rigorously connected to the theory of nuclear forces as defined, for example, by 
effective field theory \cite{drut2010}. One possibility is to derive local energy 
functionals from chiral effective field potentials by using the density matrix 
expansion \cite{stoitsov2010,carlsson2010}. Another is to use, for example, many-body 
perturbation theory to expand energy and norm kernels of ab initio approaches in a 
form amenable to DFT treatment \cite{duguet2015}; see also Duguet {\it et al}. in this issue. 

Second, irrespective of its mathematical form and physical origin, DFT kernels 
will always contain a phenomenological component in the sense that they depend on a 
small set of parameters that must be adjusted to data. In addition to making the 
theory usable, adjusting these parameters will effectively provide an ad hoc
mechanism to capture missing correlations. However, this optimization will induce 
an obvious dependence on the data and the optimization process itself, in addition 
to the pre-existing uncertainties related to the form of the functional 
and possible truncation errors in the numerical implementation. As a result of 
these uncertainties, it is by no means guaranteed that the theory will be capable 
of converging to the exact result. Therefore, 
rigorous methodology is essential in order to identify, quantify and propagate
model uncertainties.

In this paper, we review the progress made in this area over the past 10 years. 
In sect.~\ref{sec:DFT}, we recall the essential aspects of nuclear density 
functional theory; in particular we discuss the distinction between the 
self-consistent mean-field theory and the energy density functional (EDF) 
approach. In sect.~\ref{sec:DFT_model}, we look at DFT from a statistician's point 
of view: What are the parameters of the model? How can they be determined? How 
can we quantify the statistical uncertainties? We summarize the various tools used 
to answer these questions. In sect.~\ref{sec:propagation}, we review the most 
recent attempts to propagate statistical uncertainties in model predictions 
using either covariance techniques or Bayesian inference. 

%
%
%
%
\section{Nuclear Density Functional Theory}
\label{sec:DFT}

Density functional theory is a general approach for solving the quantum many-body 
problem. Its most rigorous formulation is in electronic structure theory, where 
it is based on the existence theorem of Hohenberg and Kohn \cite{hohenberg1964}. It 
states that the energy of an interacting electron gas can be written as a 
functional of the one-body local density (of electrons), and the minimum of this 
functional gives the exact ground-state of the system. Shortly thereafter, this 
formal existence theorem was supplemented with the Kohn-Sham scheme, which allows one
to determine the actual density of electrons that minimizes the energy (if the 
functional itself is known) by solving equations analogous to Hartree equations 
\cite{kohn1965}. Various extensions have been proposed to handle exchange energy 
exactly (the Kohn-Sham equations then are similar to Hartree-Fock equations), 
excited states, systems at finite temperature, and to reproduce superfluid 
correlations (see, e.g., \cite{parr1989,dreizler1990}). These extensions rely on 
reformulating the Kohn-Sham scheme with the full one-body density matrix (rather 
than the local density), density operators, a combination of one- and two-body 
densities, and so forth. Often other extensions account for relativistic effects 
\cite{eschrig1996}.

Implementations of DFT in nuclear physics are less straightforward, since the 
nuclear Hamiltonian is not known, in contrast with electronic structure 
theory. In addition, nuclei are self-bound, and correlation effects are much 
stronger than in electron systems \cite{drut2010}. Consequently, most nuclear 
energy functionals used so far have been in fact derived from the expectation 
value on the quasiparticle vacuum of effective nuclear forces used in the 
self-consistent nuclear mean-field theory \cite{bender2003}. Therefore, they 
are formulated in terms of the intrinsic one-body nonlocal density matrix and 
nonlocal pairing tensor, which can break symmetries of realistic nuclear forces 
such as translational or rotational invariance, parity, time-reversal 
invariance, and particle number. This spontaneous symmetry breaking is essential 
for introducing long-range correlations in the nuclear wavefunction 
\cite{ring2000,scheidenberger2014}. Nuclear energy functionals are, therefore, 
substantially different from their counterpart in electronic DFT. This difference
is 
reflected in the appellation of energy density functional (EDF) formalism.

\subsection{The Energy Density Functional Approach}
\label{subsec:EDF}

In this section, we succinctly describe the basic ingredients 
of the single-reference EDF (SR-EDF) approach with nonrelativistic empirical 
functionals such as derived from the Skyrme or Gogny effective interactions. We 
refer to \cite{drut2010,scheidenberger2014} for discussions of more 
general frameworks such as multireference EDF and ab initio DFT. The starting 
point is a set of single-particle states $|i\rangle$ that form a basis of the one-
body Hilbert space. The related creation/annihilation operators are 
$c_i^{\dagger}$ and $c_i$ and define the configuration space representation of the 
Fock space \cite{blaizot1985,ring2000}. The coordinate space representation is 
obtained by invoking the continuous basis $|x\rangle \equiv |
\gras{r}\sigma\tau\rangle$ of the one-body Hilbert space, with $\sigma=\pm 1/2$ 
the intrinsic spin projection and $\tau=\pm 1/2$ the isospin projection. The 
single-particle functions are then $\langle x | i \rangle = \phi(x)$, and the 
corresponding creation/annihilation operators are the field operators $c^{\dagger}
(x)$ and $c(x)$. The particle vacuum of the Fock space is denoted by $|0\rangle$
and is characterized by $\forall i,\ c_{i}|0\rangle = 0$, or, alternatively, 
$\forall x,\ c(x)|0\rangle = 0$.

Because of the importance of pairing correlations in low-energy nuclear structure 
\cite{brink2005}, we introduce a canonical transformation between particle 
operators and quasiparticle operators $\beta_{\mu}, \beta^{\dagger}_{\mu}$. This 
Bogoliubov-Valatin transformation is
\cite{valatin1961,mang1975,blaizot1985,bender2003}
\begin{equation}
\begin{array}{lll}
\beta_{\mu}           & = & \displaystyle
\sum_{m} \left[ U^{\dagger}_{\mu m}\,c_{m} +
                         V^{\dagger}_{\mu m}\,c_{m}^{\dagger} \right] ,
		     \medskip\\
\beta_{\mu}^{\dagger} & = & \displaystyle
\sum_{m} \left[ V^{T}_{\mu m}\,c_{m} +
		                U^{T}_{\mu m}\,c_{m}^{\dagger}
                     \right].
\label{eq:bogo}
\end{array}
\end{equation}
In the SR-EDF approach, we introduce the reference state $|\Phi\rangle$ as a 
product wavefunction of quasiparticle operators acting on the particle vacuum,
\begin{equation}
|\Phi\rangle = \prod_{\mu} \beta_{\mu} |0\rangle.
\label{eq:HFBvacuum}
\end{equation}
Note that, by construction, the quasiparticle vacuum (\ref{eq:HFBvacuum}) does not 
conserve the particle number. 

The next step is to recall that for any given many-body state $|\Psi\rangle$, the 
one-body density matrix $\rho$ and two-body pairing tensor $\kappa$ are defined in 
configuration space as
\begin{equation}
\rho_{ij} 
= 
\frac{\langle \Psi| c_{j}^{\dagger}c_{i} | \Psi\rangle}{\langle\Psi|\Psi\rangle},
\label{eq:rho}\ \ \ 
\kappa_{ij} 
= 
\frac{\langle \Psi| c_{j}c_{i} | \Psi\rangle}{\langle\Psi|\Psi\rangle},
\end{equation}
and the generalized density $\mathcal{R}$ as
\begin{equation}
\mathcal{R} = \left( \begin{array}{cc} \rho & \kappa \\ -\kappa^{*} & 1-\rho^{*} \end{array}\right).
\label{eq:R}
\end{equation}
When $|\Psi\rangle$ is the quasiparticle vacuum (\ref{eq:HFBvacuum}), the 
corresponding generalized density matrix verifies $\mathcal{R}^{2} = \mathcal{R}$ 
and $\mathcal{R}^{\dagger} = \mathcal{R}$ \cite{blaizot1985}. In addition, the 
Wick theorem ensures that the expectation values of any operator on the 
quasiparticle vacuum can be expressed as functions of $\rho$, $\kappa$ and 
$\kappa^{*}$ alone. These three mathematical objects are thus the basic degrees of 
freedom of the theory. In particular, the energy is then expressed as a 
functional $E[\rho,\kappa,\kappa^{*}]$.

The actual density and pairing tensor of the nucleus in its ground-state are 
determined by solving the Hartree-Fock-Bogoliubov (HFB) equations, which are obtained by applying the 
variational principle with respect to $\rho$, $\kappa$, and $\kappa^{*}$ 
\cite{blaizot1985,ring2000,bender2003}. This leads to
\begin{equation}
[ \mathcal{H}, \mathcal{R} ] = 0,
\label{eq:HFBequations}
\end{equation}
where $\mathcal{H}$ is the HFB matrix, $\mathcal{H}_{ij} = \partial E/ \partial 
\mathcal{R}_{ji}$. One-body observables can then be computed as the trace of the 
relevant operator and $\rho$. Because of the nonlinear nature of the 
HFB equations, it is possible for the generalized density to break various 
symmetries of nuclear forces. Conversely, conserved symmetries can be used to 
label quasiparticle states 
\cite{dobaczewski2000,dobaczewski2000-a,rohoziski2010}.

\subsection{Pseudopotentials and Energy Functionals}
\label{subsec:forces}

Until recently, most applications of the nuclear EDF approach have 
been based on semi-empirical EDFs explicitly derived from the expectation value of 
effective two-body forces $\hat{V}_{\text{eff.}}$ on the quasiparticle vacuum,
\begin{equation}
E[\rho,\kappa,\kappa^{*}] = \frac{\langle\Phi | \hat{T} + \hat{V}_{\text{eff.}}|\Phi\rangle}{\langle\Phi | \Phi\rangle}.
\end{equation}
In particular, the Skyrme effective force is a zero-range two-body pseudopotential 
for which the EDF becomes a functional of the local density only 
\cite{skyrme1959,vautherin1972}. The Gogny force has a finite range and gives a 
functional of the non-local one-body density \cite{decharge1980}; see, for example, 
\cite{bender2003,stone2007} for comprehensive reviews of applications of Skyrme 
and Gogny EDFs. The empirical nature of both the Skyrme and Gogny potentials is 
manifested by the presence of density dependencies, which prohibits writing the 
potential in strict second quantization form \cite{erler2010}. With the exception 
of a few recent applications 
\cite{kortelainen2010,kortelainen2012,kortelainen2014}, these EDFs have been used 
in the context of the self-consistent mean-field theory rather than in a strict 
Kohn-Sham scheme.

In particular, many applications used the underlying effective pseudopotential 
$\hat{V}_{\text{eff.}}$ to implement beyond mean-field techniques, where 
EDF reference states of the type (\ref{eq:HFBvacuum}) serve as basis states to 
expand the unknown many-body wavefunction, for example, in the generator coordinate 
method, or to restore broken symmetries by using projection techniques 
\cite{ring2000,bender2003}. A few years ago, however, 
standard beyond mean-field techniques were shown to be invalid with density-dependent 
pseudopotentials 
\cite{anguiano2001,stoitsov2007,bender2009,duguet2009,lacroix2009}. 
This result has stimulated efforts to remove density dependencies, for example by using 
momentum-dependent two-body pseudopotentials \cite{raimondi2014} or zero-range 
two- and three-body pseudopotentials \cite{sadoudi2013,sadoudi2013-a}. Since these 
pseudopotentials are specifically designed to enable beyond mean-field techniques 
such as projection and configuration mixing, the central element of all these 
approaches is the effective Hamiltonian $\hat{T} + \hat{V}_{\text{eff.}}$ rather than the EDF itself.

An alternative route is to implement a strict Kohn-Sham approach, where the only 
degrees of freedom are $\rho$, $\kappa$, and $\kappa^{*}$, the ground-state 
wave function is always a quasiparticle vacuum of the form (\ref{eq:HFBvacuum}), 
and there is no mention of some underlying effective potential 
$\hat{V}_{\text{eff.}}$. In such an approach, the energy functional 
$E[\rho,\kappa,\kappa^{*}]$ must be designed so that it contains all 
relevant types of correlations. Only two main families of such functionals have 
been proposed in the literature: those proposed by Fayans and collaborators 
\cite{fayans1994,kroemer1995,fayans2000}, and the BPCM functionals from the 
Barcelona-Paris-Cata{\~n}a-Madrid collaboration \cite{baldo2008,baldo2013}. The 
main difficulty of this strict Kohn-Sham scheme, which is more in line with the 
spirit of DFT as encountered in electronic structure theory, is to incorporate 
beyond mean-field correlations accounting, for example, for large amplitude collective 
motion, or symmetry restoration. Recent work suggests that this could be 
achieved by introducing new densities representing collective degrees of freedom 
such as two-body or ``collective'' densities 
\cite{hupin2011,hupin2011-a,hupin2012,lesinski2014} (which may lead to a 
generalization of the Kohn-Sham equations) or by adding specific terms to the 
functional designed to cancel symmetry-breaking \cite{dobaczewski2009,wang2014}. 

\subsection{Pairing Correlations}
\label{subsec:pairing}

Sometimes overlooked is the fact that, according to the Hohenberg\-Kohn theorem, the 
exact ground-state of the system can in principle be expressed entirely as a 
functional of the local one-body density matrix only. If this theorem could be 
extended directly to nuclear functionals of the intrinsic one-body density (see, 
e.g., \cite{engel2007,messud2009}), pairing correlations could --- in principle --- 
be produced by an unique functional of $\rho(\gras{r})$. In this idealized 
scenario, there would be no need for quasiparticle operators, the Bogoliubov 
transformation, or the pairing tensor: the Kohn-Sham scheme would be implemented 
directly with EDF reference states taken as particle number conserving Slater 
determinants. 

In practice, of course, the form of this functional is totally unknown. Until 
further notice, therefore, it seems more reasonable to build on the success 
of the self-consistent mean-field theory, to seek an explicit pairing term that 
is a functional of the usual pairing tensor, and to work with symmetry-breaking 
reference states of the type (\ref{eq:HFBvacuum}). Recall that the pairing tensor 
is defined from the specific form that the two-body correlation function takes in 
an HFB vacuum \cite{parr1989,hupin2011}. If we denote $\rho_{2}(x_1,x_2,x'_1,x'_2)$ 
as the full, nonlocal, two-body density matrix, then we have
\begin{multline}
\rho_{2}(x_1,x_2,x'_1,x'_2) 
=
\kappa^{*}(x_1,x_2)\kappa(x'_2,x'_1) \\
-
\rho(x'_2,x_1)\rho(x'_1,x_2) 
+ 
\rho(x'_1,x_1)\rho(x'_2,x_2).
\end{multline}
Because of this property, the pairing tensor $\kappa$ and its complex conjugate 
$\kappa^{*}$ are, indeed, the two only degrees of freedom needed to account for 
pairing correlations at the HFB level.

The pairing EDF can then be obtained by taking the expectation value of a 
pseudopotential $\hat{V}_{\text{eff.}}^{(\text{pair})}$ on the quasiparticle 
vacuum, which will immediately introduce a dependence on $\kappa^{*}$ and 
$\kappa$. This potential can be the same as the one used in the particle-hole 
channel, which is typically the choice retained when working with the Gogny force 
\cite{decharge1980}. It can also have a different form, ranging from  
simple seniority pairing forces \cite{ring2000} to density-dependent zero-range 
pairing forces \cite{dobaczewski2002} to separable expansion of finite-range, 
Gogny-like potentials \cite{tian2009,tian2009-a}. Most of these pairing forces, 
and hence the resulting pairing functionals, are characterized by only a few parameters, 
and all lead to EDFs that are functionals of $\kappa^*$ and $\kappa$ only. 

%
%
%
%
\section{Density Functional Theory as a Model}
\label{sec:DFT_model}

Whether building the description of atomic nuclei on an effective potential 
$\hat{V}_{\text{eff.}}$ that defines both the mean-field and beyond mean-field 
corrections, or an EDF $E[\rho,\kappa,\kappa^{*}]$ in a strict Kohn-Sham 
framework, theoretical predictions will depend on a set of unknown parameters 
$\gras{x}$ corresponding, respectively, to the parameters of the effective nuclear 
force or the coupling constants of the EDF. Some of these parameters may be 
constrained by exploring the connections with the theory of realistic nuclear 
forces or investigating ideal systems such as nuclear matter or neutron drops 
\cite{bogner2011,gandolfi2011}. In general, however, one will also have to 
introduce experimental data in nuclei in order to set the values of these parameters. This 
fit of low-energy coupling constants to experimental data belongs to the class of 
inverse problems in statistics. In this section, we review some of the techniques 
used in nuclear DFT to solve this problem. Most of our considerations are 
based on the SR-EDF approach to nuclear structure but are easily extended to the 
self-consistent mean-field approach.

\subsection{Parameter Estimation}
\label{subsec:optimization}

The problem of determining the parameters of the nuclear EDF is easily posed: 
one needs only to choose a set of data points, define an objective function such as a $\chi^2$ function, and minimize the objective function with respect to 
the parameters. We use the following notations: $\gras{y}$ 
denotes the values of a set of experimental observables, with $y_{ti}$ the value 
of the $i$th observable of type $t$; $\gras{x} \equiv (x_{1}, 
\dots,x_{n_{x}})$ represent the vector of the $n_{x}$ parameters of the model, 
that is, the EDF coupling constants in our case; and $\gras{\eta}$ collects the 
output of all model calculations. In our case, $\eta_{ti}(\gras{x})$ is thus the
output of an HFB calculation for the $i$th observable of type $t$. 
Also, $\gras{\epsilon}$ is the vector containing the error between the actual 
calculation and the experimental value. By definition, we thus have
\begin{equation}
y_{ti} = \eta_{ti}(\gras{x}) + \epsilon_{ti},\ 
\epsilon_{ti} \stackrel{\rm indep}{\sim} N(0,\sigma_t),
\ \ \forall (t,i).
\end{equation}
Based on these notations, we minimize the weighted mean squared 
deviation given by
\begin{equation}
\chi^2(\gras{x}) = \frac{1}{n_{d} - n_{x}} \sum_{t=1}^{T} \sum_{i=1}^{n_{t}}
\left( \frac{y_{ti}-\eta_{ti}(\gras{x})}{\sigma_t} \right)^{2},
\label{eq:chi2}
\end{equation}
where $T$ is the total number of different data types, $n_{t}$ the total 
number of points of type $t$, and $n_{d}$ the total number of experimental 
points, $n_{d} = \sum_{t=1}^{T} n_{t}$. By convention, the vector of parameters 
at the minimum of the $\chi^2$ is noted $\hat{\gras{x}}$. Recall that 
$\chi^2 \gg 1$ implies a poor fit, where $\chi^2 \approx 1$ indicates a good 
fit. This is the familiar ``$\chi^2$ per degree of freedom.'' If all 
errors $\epsilon_{ti}$ are independent and normally distributed with 
mean 0, then the minimization of (\ref{eq:chi2}) is equivalent to maximizing 
the likelihood function \cite{metzger2002,brandt2014}. In addition, the 
$\chi^2$ is a random variable that follows a genuine $\chi^2$ probability 
distribution function.

\subsubsection{Experimental Dataset and Bias Estimation}
\label{subsubsec:data}

Choosing which and how many data points to include in the $\chi^2$ is the 
first important decision, and several strategies have been followed. In nuclear 
mass models based either on the Skyrme or Gogny force, all available experimental 
information on atomic masses is used; see \cite{goriely2009,goriely2013} 
and references therein. It is supplemented by additional data on, for example, fission 
barriers \cite{goriely2007} or neutron matter \cite{goriely2013}. The main 
concern for mass models is the risk of producing a high-bias estimator of the 
data. In simpler terms, it is by no means guaranteed that mass model 
parameterizations of Skyrme of Gogny forces are reliable for computing 
observables that are not masses.

By contrast, most historical fits of the Skyrme and Gogny forces were based on 
the smallest possible set of data. These included nuclear matter properties, 
binding energies, radii, and single-particle states; see \cite{bender2003,stone2007,kluepfel2009} 
for a discussion. In addition, data in finite nuclei was taken almost exclusively 
in doubly-magic spherical nuclei. Two of the most notable exceptions are the 
SkM* parameterization of the Skyrme force \cite{bartel1982} and the D1S 
parameterization of the Gogny force \cite{berger1991}, which included information 
on the fission barrier in $^{240}$Pu. The combination of small dataset and 
spherical nuclei is also likely to lead to high-bias estimators.

\begin{figure}[!ht]
\center
\includegraphics[width=0.95\linewidth]{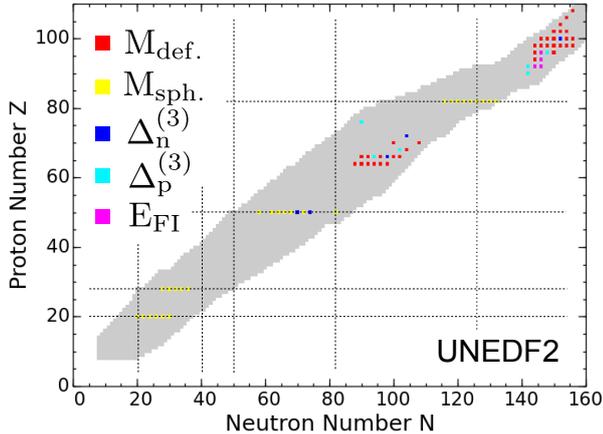}
\caption{Experimental dataset used for optimizing the UNEDF2 Skyrme 
functional \cite{kortelainen2014}.}
\label{fig:data}
\end{figure}

The recently proposed UNEDF parameterizations of the Skyrme EDF represent an 
attempt to reduce the bias of the fitting procedure by selecting a medium-sized 
sample of 100+ data points carefully selected in both spherical and deformed 
nuclei; see fig.~\ref{fig:data} for the specific case of the UNEDF2 functional 
\cite{kortelainen2010,kortelainen2012,kortelainen2014}. As a result, the ability 
of UNEDF functionals to reproduce masses or fission barriers has degraded between 
UNEDF0 (3 datatypes, 108 points) and UNEDF2 (5 data types, 130 points), while the 
predictive power on binding energies near closed shells and on single-particle 
states increased. Note that until now, no attempt has been made to rigorously 
quantify the bias of EDF parameterizations.

The pairing channel represents an additional difficulty when determining the 
parameters of the nuclear EDF. Indeed, very little data can 
effectively and unambiguously constrain the pairing functional directly at the 
HFB level. In practice, the odd-even staggering of binding energies is most often 
used \cite{satula1998,rutz1999,dobaczewski2001,duguet2001,bertsch2009}. The 
UNEDF functionals were the first ones where the fit of the pairing functional was 
performed simultaneously with the fit of the Skyrme EDF. As a result, there 
are built-in correlations between the parameters of the Skyrme EDF and the two 
parameters that control the pairing functional.

\subsubsection{Optimization Algorithm}
\label{subsubsec:algorithm}

The minimization of the $\chi^2$ function in the context of nuclear DFT remains 
costly in computational resources. In the example of the UNEDF 
functionals, 100+ full, axially-deformed HFB calculations must be performed 
in order to 
define the $\chi^2$. Some of the most popular parameterizations of the nuclear 
EDF were published in the 1980s and 1990s, where the cost of running a full 
HFB calculation was prohibitive in terms of $\chi^2$ minimization. 
Even now, the optimization of HFB mass models, where the $\chi^2$ includes over 
2,500 points, is still performed in spherical symmetry by using empirical 
renormalizations \cite{tondeur2000}. 

In view of this computational cost, specifically designed algorithms with a 
focus on efficiency and robustness are especially valuable. We recall that 
derivatives $\partial\eta_{ti}(\gras{x})/\partial x_{\mu} $ are not available 
analytically for the minimization of the $\chi^2$ (\ref{eq:chi2}). Of course 
they can always be computed numerically, but at a significant cost when $n_{x}$ 
is large. Therefore, the optimization of the nuclear EDF is most efficiently 
performed with derivative-free approaches.

\begin{figure}[!ht]
\center
\includegraphics[width=0.95\linewidth]{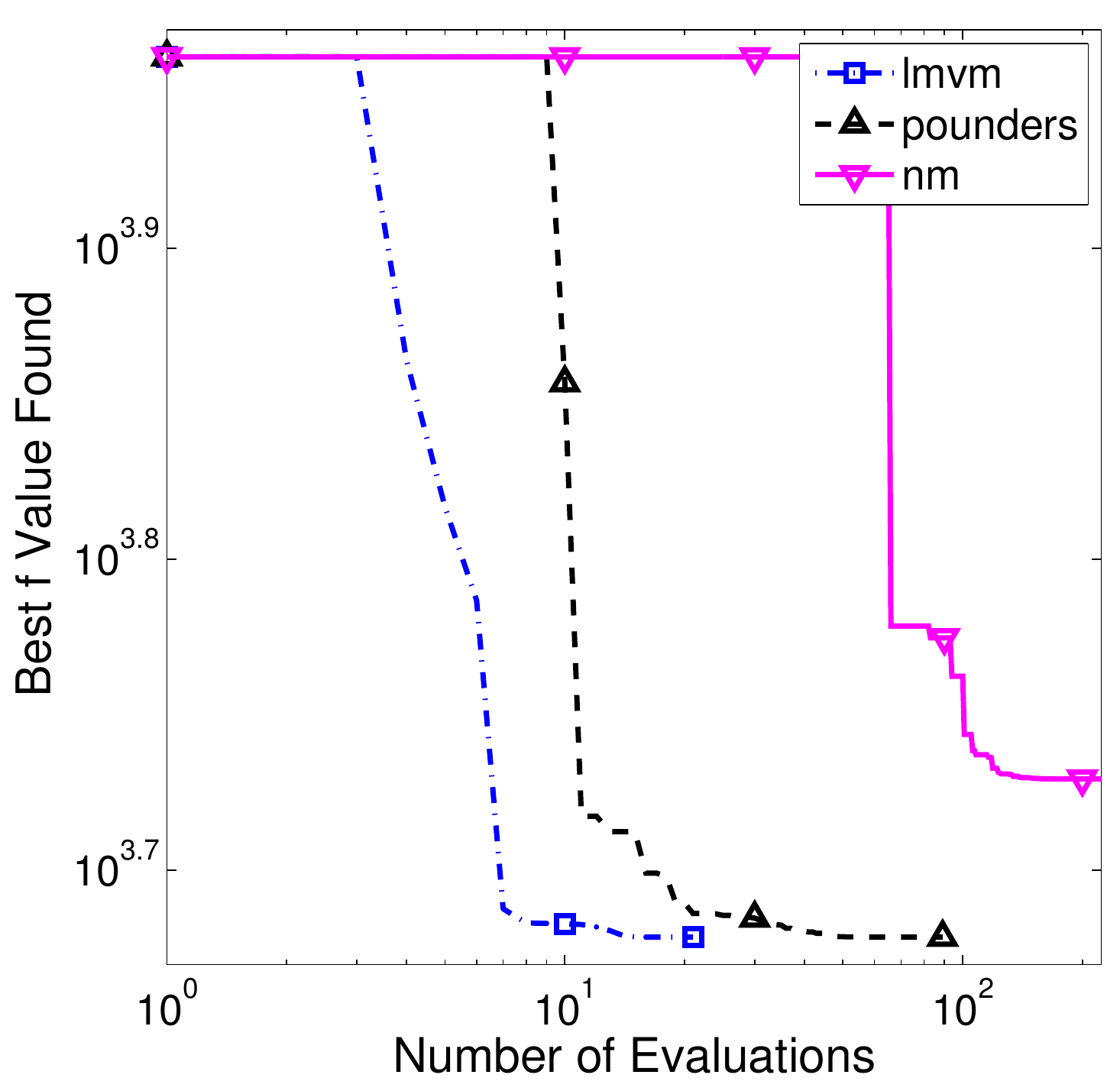}
\caption{Performance (log-log scale) of three solvers (limited-memory variable 
metric, \pounders\, Nelder-Mead) for nonlinear generalized $\chi^2$ problems 
with $n_{x}=6$, $n_{d}=428$; from \cite{wild2015}.}
\label{fig:algo}
\end{figure}

Although a rich literature on the subject exists, we mention here only the 
\pounders\  algorithm developed in the framework of the UNEDF collaboration 
\cite{kortelainen2010,tao-man,SWCHAP14,wild2015}. \pounders\ is a 
derivative-free trust-region method based on forming a local 
quadratic model of each component of the $\chi^2$. The 
quadratic models are valid only in a small region of the parameter space near 
the 
current point $\gras{x}$, but their aggregate approximation can be minimized 
analytically. The minimum 
$\gras{x}+\delta\gras{x}$ defines the next point of a Newton-like procedure. As 
seen in fig.~\ref{fig:algo}, the \pounders\  algorithm converges quickly 
compared to the traditional Nelder-Mead algorithm; most important, it gives a 
better solution.
\pounders\  has recently been applied to other problems of interest in nuclear 
physics \cite{ekstroem2013,bertolli2012}.

\subsubsection{Choice of the Objective Function}
\label{subsubsec:function}

\begin{table}[-ht]
\center
\caption{Root-mean-square deviations for each observable in the UNEDF1
optimization protocol compared for UNEDF1-HFB for a few different values of
the standard deviation $\sigma_{OES}$ (in MeV) of the OES data. All r.m.s.\ 
values are in MeV except the ones for proton radii, which are in fm; from 
\cite{schunck2015}.}
\begin{tabular}{@{}ccccc}
\hline
 $\sigma_{OES}$   & $0.025$ & $0.050$ &
                  $0.075$  &  $0.100$ \\
\hline
 Deformed masses  & 0.944 &  0.776 & 2.596 &  0.806 \\
 Spherical masses & 2.427 &  1.836 & 2.669 &  1.718 \\
 Proton radii     & 0.022 &  0.022 & 0.022 &  0.022 \\
 OES neutrons     & 0.012 &  0.051 & 0.065 &  0.080 \\
 OES protons      & 0.043 &  0.074 & 0.075 &  0.072 \\
 Fission isomer   & 0.809 &  0.558 & 0.535 &  0.530 \\
\hline
\end{tabular}
\label{tab:weights}
\end{table}

\begin{table*}[tb]
\caption{Rerun of \pounders\ on the UNEDF0 problem ($n_d=108$, $n_N=72$) from 
two different starting points. The scaled difference columns represent the 
difference between the final value found and the original UNEDF0 parameterization, 
scaled by its uncertainties $\sigma_i$; see table VIII in \cite{kortelainen2010}.}
\begin{center}
\begin{tabular}{r||llr|llr}
& \multicolumn{2}{c}{\textbf{Starting from SLy4}} & \textbf{Scaled Diff.}
& \multicolumn{2}{|c}{\textbf{Starting from SKM*}} & \textbf{Scaled Diff.} \\ 
& \textit{initial} & \textit{final} & & \textit{initial} & \textit{final} \\ 
\hline
$\rho_c$ &     0.159539 & \same{0.160}486 &   -0.03954 &     0.160319 & 
   \same{0.160}435 &   -0.09106 \\ 
$E^{NM}/A$ &     -15.9721 &     \same{-16.0}685 &     -0.2285 &          -16 &  
   \same{-16.0}73 &    -0.3119 \\ 
$K^{NM}$ &      229.901 &          \underline{\same{230}} &  \textendash &      216.658 &          \underline{\same{230}} &            \textendash \\ 
$a^{NM}_{\mathrm{sym}}$ &      32.0043 &      \same{3}1.3393 &     0.2604 &    
 30.0324 &      \same{3}1.7221 &     0.3856 \\ 
$L^{NM}_{\mathrm{sym}}$ &      45.9618 &      54.2493 &     0.2290 &     
45.7704 &      60.4725 &     0.3844 \\ 
$1/,M_s^*$ &      1.43955 &          \underline{\same{0.9}} &   \textendash  &      1.26826 &          \underline{\same{0.9}} &            \textendash \\ 
$C^{\rho \Delta \rho}_0$ &     -76.9962 &     \same{-55.2}344 &    0.01545 &    
-68.2031 &     \same{-55}.7348 &    -0.2794 \\ 
$C^{\rho \Delta \rho}_1$ &      15.6571 &     -64.1619 &    -0.1499 &     
17.1094 &     -70.4274 &    -0.2599 \\ 
$V^n_0$ &      -285.84 &     \same{-170}.796 &    -0.2003 &         -280 &    
\same{-170}.788 &    -0.1966 \\ 
$V^p_0$ &      -285.84 &     \same{-19}7.782 &     0.4238 &         -280 &    
\same{-19}8.038 &     0.3474 \\ 
$C^{\rho \nabla J}_0$ &       -92.25 &     \same{-7}7.9436 &      0.4637 &      
 -97.5 &     \same{-79}.2915 &    0.06990 \\ 
$C^{\rho \nabla J}_1$ &       -30.75 &      27.4519 &    -0.6171 &        -32.5
&      \same{4}9.5737 &      0.1339 \\ 
\hline 
$f(\xh)$ & 1188.75 &      \same{67}.9034 & & 24814.1 &      \same{67}.5738 \\ 
$n_f$ & & 235 & & & 150 \\ 
\end{tabular}
\label{tab:starting}
\end{center}
\end{table*}

Given a set of data points and a minimization algorithm, some 
latitude remains in defining the weights, or standard deviations $\sigma_{t}$, associated 
with each data type. These quantities represent the estimated error on the 
data type $t$. They are in principle determined in such a way that the $\chi^2$ 
objective function (\ref{eq:chi2}) approaches 1 at the minimum. Satisfying this 
condition may require readjusting the weights during the minimization 
\cite{dobaczewski2014}. In practice, this step has rarely been done in nuclear EDF 
optimization, and the weights are most often kept constant (though data type 
dependent). Results reported in \cite{schunck2015} and listed in table 
\ref{tab:weights} show that the impact of the weights on the result of the 
optimization could be significant.

Minimizing the $\chi^2$ (\ref{eq:chi2}) requires initializing the optimization 
algorithm with a vector $\gras{x}_{0}$. Ideally, the optimization algorithm would 
be able to converge to the absolute minimum of the objective function, given 
a set of constraints dictated by reality. In practice, it is nearly 
impossible to guarantee such a result. To our knowledge, there is only one 
example where the impact of the initial point on the resulting parameterization 
was studied in detail \cite{wild2015}. Table \ref{tab:starting} 
illustrates the robustness of the \pounders\ algorithm; similar solutions to 
the optimization problem are obtained when starting from the SLy4 or SkM* 
parameterization. The largest difference occurs for $L^{NM}_{\mathrm{sym}}$ (slope 
of the symmetry energy in nuclear matter at saturation density) and $C^{\rho 
\nabla J}_1$ (isovector spin-orbit coupling constant), which are poorly 
constrained by the dataset.

\subsection{Statistical Uncertainties of Energy Densities}
\label{subsec:uq}

Irrespective of the form of the energy functional, the degree of arbitrariness in 
defining the $\chi^2$ used to determine the ``best'' parameters of the EDF 
clearly suggests possibly large uncertainties in the resulting parameterizations. 
Standard methods of probability and statistics can be used to quantify some of 
these uncertainties. In this section, we review only the techniques used to 
estimate statistical uncertainties. Few studies of systematic 
uncertainties have been conducted so far; see \cite{schunck2015} for discussion. Numerical 
errors are discussed separately in sect.~\ref{subsec:numerics}. This separation 
is made for convenience only, since it is illusory to think that all sources of 
uncertainties can be completely disentangled.

\subsubsection{Covariance Analysis}
\label{subsubsec:covariance}

One of the most common quantities used for estimating statistical uncertainties is 
the covariance matrix. The use of covariance techniques in nuclear DFT is 
relatively recent. Full regression analysis was first introduced in the context 
of nuclear mass fits in \cite{toivanen2008}. The covariance matrix was first 
mentioned and computed for Skyrme EDF optimization in \cite{kluepfel2009}. Since 
then, there have been many applications of this technique to compute the 
standard deviation of EDF parameters and propagate uncertainties in model 
predictions; see sect.~\ref{sec:propagation}.

In the following, we denote $\gras{C}_{M}$ the covariance matrix of the 
parameters $\gras{x}$ of the model $M$ that we are using (EDF), formally,
\begin{equation}
(C_M)_{ij} = E\left[ (x_i - E(x_i)) (x_j - E(x_j)) \right],
\end{equation}
where $E()$ refers to the average of a random variable; each parameter $x_i$ 
is thus treated as a random variable. One should distinguish $\gras{C}_{M}$ from 
the ``data'' covariance matrix $\gras{C}_{D}$. The latter notation will be used 
to refer to the covariance matrix of the random variables $\gras{\epsilon}$ 
associated with the error between the model output $\gras{\eta}(\gras{x})$ and 
the experimental data $\gras{y}$. All these 
misfits $\epsilon_{ti}$ are often assumed to be independent, therefore $\gras{C}_{D}$ is diagonal and  
$(C_D)_{ij} = \sigma_i^{2}\delta_{ij}$. One also assumes that they follow a 
(multivariate) normal distribution with mean 0, 
$\gras{\epsilon} \sim \mathcal{N} (0, \gras{C}_D)$. 

In the simple case of an unweighted, linear least-squares optimization, where 
$\gras{\eta}(\gras{x}) = \gras{A}\gras{x}$ and $\sigma_i^2=1$, the inverse of 
the covariance matrix 
$\gras{C}_{M}$ can be computed as \cite{metzger2002,brandt2014}
\begin{equation}
(C_{M}^{-1})_{ij}(\gras{x}) = 
\left( \frac{1}{2} \frac{\partial^{2}\chi^2}{\partial x_i \partial x_j} 
\right)^{-1}
= 
2\left(\mathcal{H}^{-1}\right)_{ij},
\end{equation}
where $\mathcal{H}$ is the Hessian matrix of $(n_d-n_x)\chi^2(\gras{x})$. In the case 
of nuclear EDF optimization, the quality of the covariance matrix estimation is 
thus contingent on the linear dependence of observables with model parameters 
within the range of variation of interest. From the literature, one finds that 
nuclear binding energies behave linearly across a broad range of parameter 
space \cite{schunck2015}; single-particle orbitals have a small degree of 
nonlinearity \cite{kortelainen2008}; nonlinearity becomes more pronounced in the 
variation of fission isomer excitation energies \cite{schunck2015}. While 
covariance techniques have been often employed recently to obtain estimates of 
statistical uncertainties on model predictions, see 
sect.~\ref{sec:propagation}, the underlying hypothesis of linearity has 
rarely been investigated in detail.

The covariance matrix can also be used to get an estimate of confidence 
intervals/regions. Recall that if the errors $\gras{\epsilon}$ follow a 
multivariate 
normal distribution, then the confidence interval at $\alpha$ percent for 
parameter $i$ is defined by the endpoints
\begin{equation}
\hat{x}_{i} \pm \sqrt{(C_M)_{ii}}t_{n_{d}-n_{x},1-\frac{\alpha}{2}},
\end{equation}
where $t_{n_{d}-n_{x},1-\frac{\alpha}{2}}$ is the $1 - \frac{\alpha}{2}$ 
quantile of the (Student's) $t$ distribution with $n_{d}-n_{x}$ degrees of 
freedom 
\cite{montgomery2002,metzger2002,toivanen2008,dudek2011,szpak2011}. This 
was used, for example, in the assessment of the UNEDF functionals 
\cite{kortelainen2010,kortelainen2012,kortelainen2014}. The diagonal 
elements of the covariance matrix define the standard deviations, $(C_M)_{ii}$, 
 of each parameter $i$.

%
\subsubsection{Bayesian Techniques}
\label{subsubsec:bayesian}

Bayesian inference techniques have been used for many years in the nuclear data 
community \cite{froehner2000,leeb2008,capote2008,herman2011,talou2015}. In 
nuclear structure, this method has recently gained ground, for example, to quantify 
uncertainties in chiral effective potentials \cite{schindler2009,furnstahl2015}. 
In DFT, Bayesian inference has been used in electronic structure theory to 
evaluate uncertainties induced by the fit of the exchange-correlation functional 
in the generalized gradient approximation \cite{mortensen2005}. 

Following  \cite{froehner2000}, one may describe Bayesian techniques as an 
exercise of inductive inference when the probability for an hypothesis $A$ to 
be true is interpreted not strictly as the number of observations of $A$ over 
the total number of outcomes, but rather as the degree of plausibility that $A$ 
is true. 
The ``philosophical'' interpretation is that the true value of the 
model parameters $\gras{x}$ is described probabilistically.  Additional
data further constrain the probability distribution but never reduces it
to a single, known value.

The Bayes theorem provides the mathematical foundation for Bayesian techniques. 
In the case of continuous random variables, it reads
\begin{equation}
p(\gras{x}|\gras{y},M)d\gras{x} 
= 
\frac{P(\gras{y}|\gras{x},M)P(\gras{x}|M)d\gras{x}}{\displaystyle\int P(\gras{y}|M)d\gras{x}}.
\label{eq:bayes}
\end{equation}
In practice, we seek the probability of the model $M$ having parameters 
$\gras{x}$ based on a set of observed data $\gras{y}$. The model is typically 
characterized by a number of features that add to the resulting 
uncertainties. In the case of EDF optimization, these features include the type of 
functional (Skyrme, Gogny, or other), the treatment of pairing correlations (HFB 
approximation, particle number projection), and the numerical implementation. 
The probability distribution $p(\gras{x}|\gras{y},M)$ is the {\em posterior} 
distribution. Note that the posterior is computed within the model $M$: it does 
not contain any information about the validity of said model. In other words, 
suppose the posterior distribution is sharply peaked around a given value 
$\gras{x}_{0}$: the fact that $\gras{x}_{0}$ is the most likely parameter set does 
not mean (i) that it is the correct one (since more data may change the 
distribution), and (ii) that the resulting model is the correct one (since 
everything is model--dependent).

In eq.~(\ref{eq:bayes}), $P(\gras{y}|\gras{x},M)$ is the probability that the 
model produces the data given the parameters: it is the likelihood function 
\cite{metzger2002}. $P(\gras{x}|M)$ is the probability that the model has 
parameters $\gras{x}$ irrespective of any data: it is the {\em prior} 
distribution. For a uniform prior distribution, maximizing the posterior 
distribution is equivalent to maximizing the likelihood. 
We remark in passing 
that both statistical approaches give different results if one looks at the probability 
distribution of some new parameter $g(\gras{x})$ that is a function of the 
original parameters $\gras{x}$ \cite{metzger2002}.

Bayesian inference can also be used to compute an estimate of the covariance 
matrix $C_{M}$ between the parameters. Assuming weak nonlinearities of the 
model parameters, that is, $\gras{\eta}(\gras{x}) \propto \gras{x}$, the 
likelihood function is approximately Gaussian with respect to $\gras{x}$. If 
one assumes, for simplicity, full ignorance about the prior distribution (uniform 
distribution with independent parameters), then the posterior covariance matrix 
is given by \cite{tarantola2005}
\begin{equation}
\tilde{\gras{C}}_{M}^{-1} \approx \gras{G}^{T}\gras{C}_{D}^{-1}\gras{G}, 
\end{equation}
where
\begin{equation}
G_{ij}(\gras{x}) = \frac{\partial \eta_{i}}{\partial x_{j}}(\gras{x})
\end{equation}
and $\gras{C}_{D}$ is, as before, the covariance matrix associated with the 
misfits between data and the predictions. Owing to the (near) linearity of the 
model parameters, one can easily find that $\tilde{\gras{C}}_{M}^{-1} = 
2\gras{\mathcal{H}}$ as obtained from the standard covariance matrix. The 
advantage of the Bayesian approach is the possiblity of including in the 
calculation of the covariance matrix the effect of prior knowledge of the 
distribution of model parameters; see sect.~3.2.3 in \cite{tarantola2005} for 
details.

Posterior distributions are typically sampled by using Markov chain Monte Carlo 
(MCMC) techniques \cite{gamerman2006}, the result being a (dependent) 
sequence of samples $\{\gras{x}^{(1)},\ldots,\gras{x}^{(T)}\}$. In practice, 
this sampling can be computationally challenging, since thousands or millions of 
evaluations of the likelihood function, hence of the $\chi^2$ function, may be 
needed. As mentioned in sect.~\ref{subsubsec:algorithm}, the $\chi^2$ functions 
used in nuclear EDF optimizations may typically involve between 100 and 2,500 
HFB calculations or more, making the direct sampling of the posterior 
distribution prohibitive. 
The alternative is to estimate response surface functions in order to 
emulate the behavior of the model response $\eta_{ti}(\gras{x})$ at a much cheaper cost 
\cite{kennedy2001,higdon2008,bilionis2013}. 
The parameters of these response 
functions can also be incorporated in the statistical setting $\gras{x}$.

\begin{figure}[!ht]
\center
\includegraphics[width=0.95\linewidth]{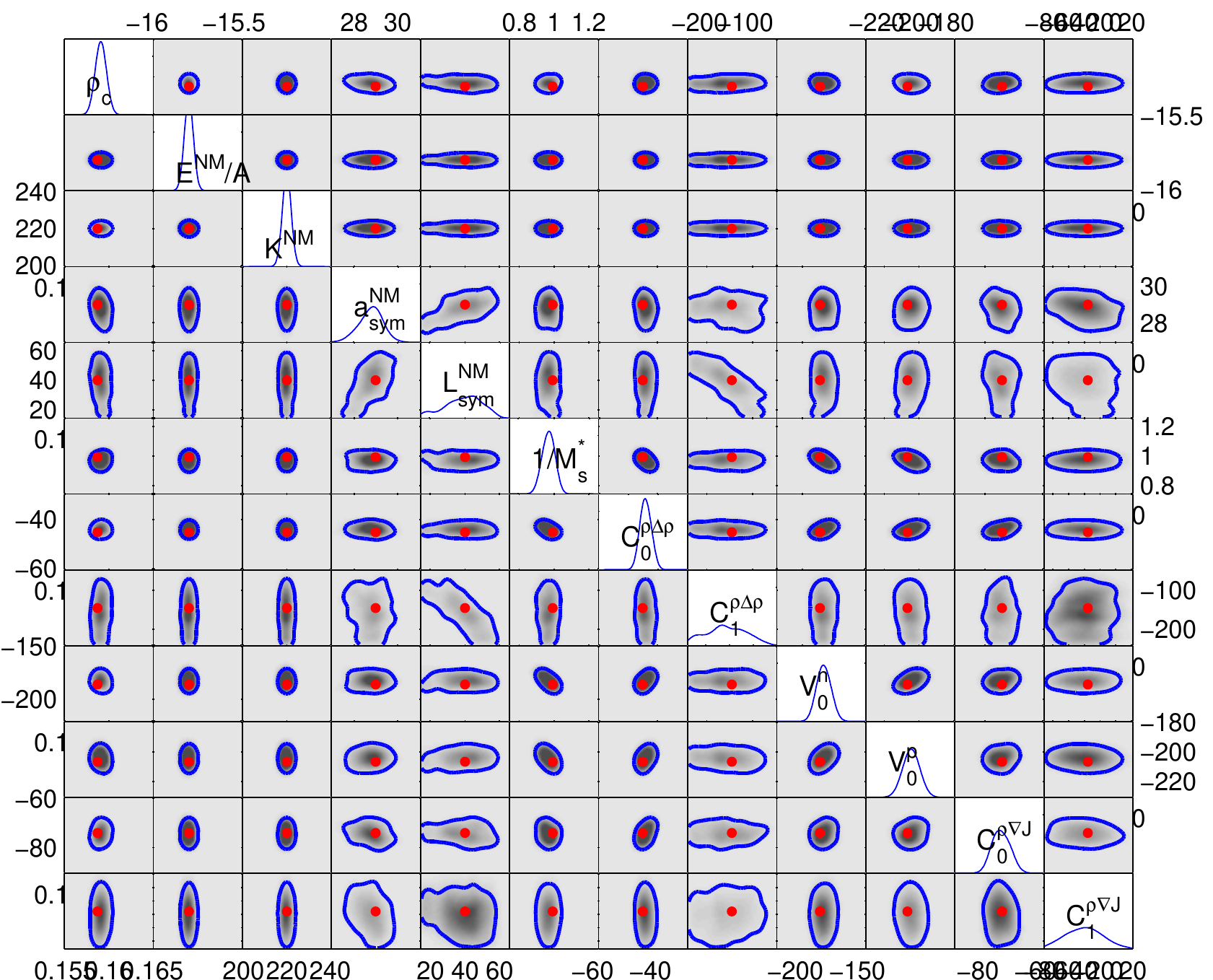}
\caption{(Color online) Univariate and bivariate marginal estimates of the 
posterior distribution for the 12-dimensional DFT parameter vector of the UNEDF1 
parameterization. The blue lines enclose an estimated 95\% region for the 
posterior distribution found when only the deformed masses from the original 
UNEDF1 data are accounted for. The red dot corresponds to the UNEDF1 values; see 
\cite{schunck2015}.}
\label{fig:posterior}
\end{figure}

Until now, there have been only two examples of Bayesian applications in nuclear 
EDF optimization. In \cite{goriely2014}, the backward forward Monte-Carlo method 
was applied to estimate uncertainties in Skyrme mass model parameters. In 
\cite{mcdonnell2015}, the full posterior distribution of the Skyrme EDF 
corresponding to the UNEDF1 $\chi^2$ was computed by using response functions based 
on Gaussian processes. The resulting 12-dimensional multivariate distribution is 
shown in fig.~\ref{fig:posterior}. The characteristics of the posterior  
distribution, such as the calculated standard deviations of the parameters, 
are similar to the results from the analysis based on confidence interval 
given in \cite{kortelainen2012}. 

\subsection{Numerical Implementations}
\label{subsec:numerics}

Implementing the DFT equation (e.g., the HFB equations for the SR-EDF approach)
in a computer program introduces numerical errors. These errors are unavoidable 
becaues the density matrix and pairing tensor have an infinite number of degrees 
of freedom. In this section, we focus only on the problem of solving the HFB 
equations: in the SR-EDF approach, these are the only equations needed. In 
multireference EDF, the HFB equations also play a central role because errors in 
the solutions will propagate to the calculation of beyond mean-field 
corrections such as in the GCM \cite{bender2003}. 

\begin{figure}[!ht]
\center
\includegraphics[width=0.95\linewidth]{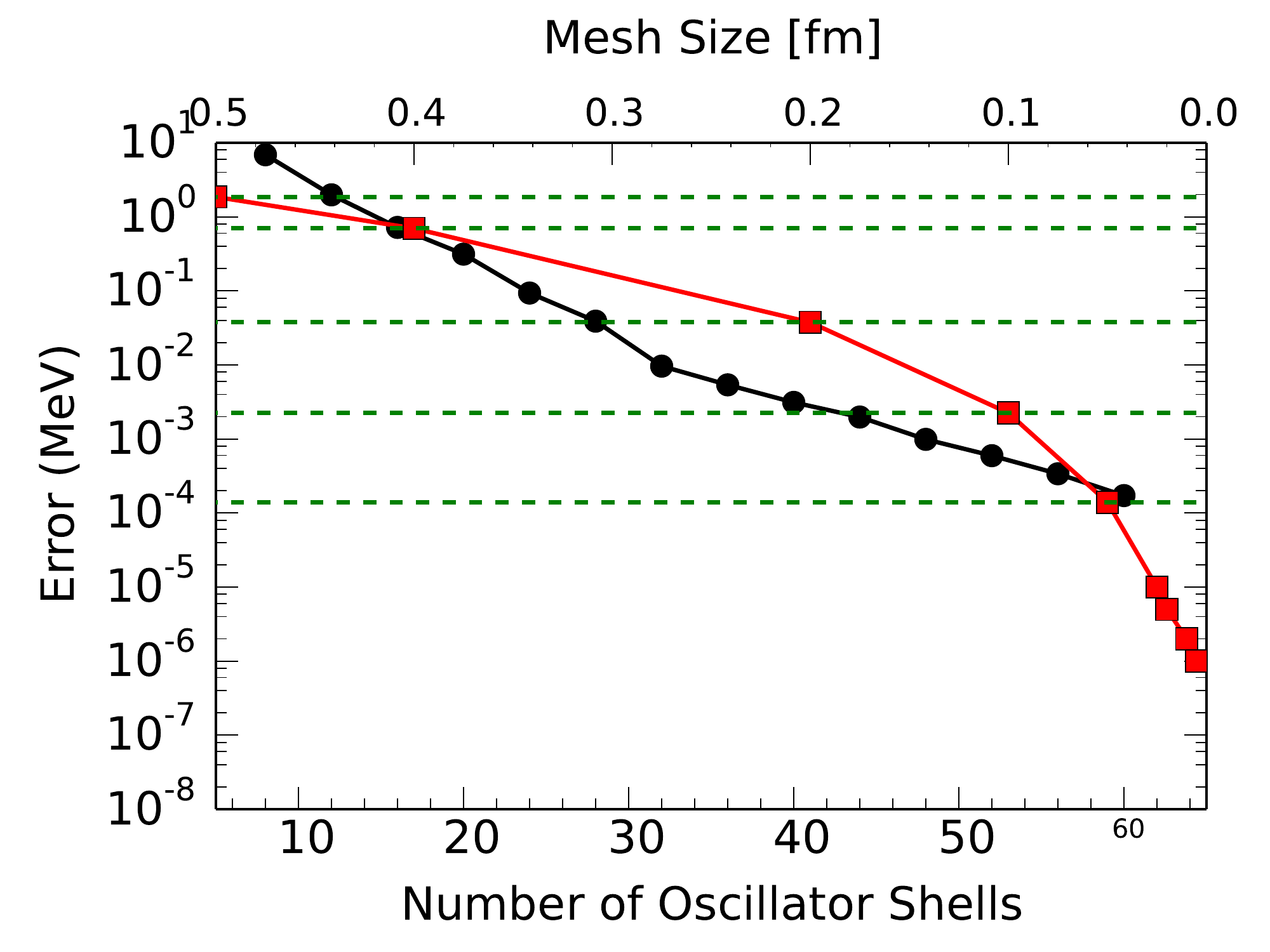}
\caption{Comparison between the pace of convergence of a spherical DFT 
calculation in coordinate-space, (red squares), and configuration space (HO basis), 
(black circles). Results were obtained by setting both direct and exchange terms of 
the Coulomb potentials to 0. The HO basis results are optimized with respect to 
the oscillator frequency. Coordinate space calculations were performed with HFBRAD
in a box of 20 fm \cite{bennaceur2005}, HO calculations with HOSPHE
\cite{carlsson2010}; from \cite{schunck2015}.}
\label{fig:convergence}
\end{figure}

One of the most popular approaches to solve the HFB equations is to expand the 
HFB solutions on a basis of known functions. In atomic nuclei, the eigenstates 
of the harmonic oscillator (HO) are most often used, since they are given 
analytically on spherical, cylindrical, and Cartesian coordinates. In addition, 
there is an exact separation between center of mass and relative motion. 
The lower part of the nuclear mean-field is well approximated by an HO. 
Several DFT solvers using HO basis expansions have been published; see 
\cite{dobaczewski1997,dobaczewski1997-a,dobaczewski2000,dobaczewski2004,dobaczewski2005,stoitsov2005,dobaczewski2009,carlsson2010,schunck2012,stoitsov2013}. 
In practice, all basis expansions are 
truncated. Therefore, HFB solutions become dependent on the characteristic 
parameters of the basis; in the case of the HO, these are the basis frequencies 
$\gras{\omega} = (\omega_x, \omega_y, \omega_z)$, number of oscillator shells 
$N$, and total number of basis states (if the basis is spherical, all frequencies 
are identical and there is a one-to-one correspondence between number of shells 
and number of states). This spurious dependence may induce large errors, for 
example in nuclei with large elongations or weakly bound systems.
\cite{schunck2013,schunck2015}.

The HFB equations can also be solved by direct numerical integration; see, for example, 
\cite{dobaczewski1984} for the HFB formalism in coordinate space. This has been 
done in spherical and axial symmetry only \cite{bennaceur2005,pei2008}. For 
more complex geometries, the computational cost of direct integration becomes 
prohibitive; and hybrid strategies such as lattice discretization 
\cite{bonche2005,ryssens2015}, finite element analysis 
\cite{poschl1997,poschl1997-a} and multi-resolution wavelet expansion 
\cite{pei2014} have been investigated. While numerically significantly more 
precise (see fig.~\ref{fig:convergence} for a comparison between 
the pace of convergence of the two methods), $r$-space-based techniques come with 
a higher computational cost, in terms of processes, memory, or disk space. 
Such approaches are also not ideal for handling finite-range local forces or 
nonlocal forces.

Numerical errors inherent in DFT solvers are often overlooked, even though they 
may play a non-negligible role in the estimation of statistical uncertainties. 
For example, the truncation error of HO expansions increases with nuclear 
deformation, even when one tries to adjust accordingly the geometry of the HO 
basis \cite{samyn2005,schunck2013}. As a result, the numerical error in the 
energy of, say, the fission isomer or the top of the fission barriers in actinide 
nuclei is always going to be larger than the error in the ground state. In 
fact, at very large deformations, the error of one-center basis expansions can 
reach a few MeV. Apart from adopting empirical corrections based on auxiliary 
large-scale surveys of numerical errors \cite{hilaire2007}, the solution could be 
to generalize asymptotic formulas such as proposed in the context of ab initio 
theory \cite{coon2012,furnstahl2012,more2013}. This problem, as well as the 
inclusion of these errors in the calculation of uncertainties, remains open.

%
%
%
%
\section{Uncertainty Propagation and Predictive Power}
\label{sec:propagation}

One of the main advantages of using the statistical analysis techniques briefly 
presented in sect.~\ref{sec:DFT_model} is to provide a rigorous framework for 
propagating the quantified uncertainties to predictions. These predictions can be 
the result of running the same model on a different dataset; for example, 
computing masses of exotic neutron-rich nuclei or superheavy elements that have 
not been included in the dataset during the optimization \cite{mcdonnell2015}.

Most important, uncertainties in the EDF could also, in principle, be 
propagated to cases where the EDF is only one of several theoretical components, 
each with a few sources of uncertainties. The calculation of low-lying 
excited states within the quasiparticle random phase approximation (QRPA) is a 
straightforward example: it typically contains approximation of its own (symmetry 
restrictions, limited model space, etc.), but it is also strongly dependent on 
the EDF.

Let us firmly reassert here that in both cases, propagating uncertainties 
estimated using covariance of Bayesian techniques provides information only about the impact 
of said uncertainties. The procedure does little to provide ways to reduce 
them. In EDF optimization, numerical errors due to basis or mesh truncation can 
easily (at least in principle) be remedied. Statistical and {\it a fortiori} 
systematic uncertainties are much more difficult to address without a detailed 
understanding of the nuclear many-body problem.

Most uncertainty propagation reported in the literature was performed with 
covariance techniques. This situation implies that computed observables are 
linearly dependent on model parameters, which is guaranteed only locally near 
the optimal point. The computed value $\eta_y(\gras{x})$ of a single new 
observable $y$ depends on the parameterization of the EDF, and one can estimate 
its standard deviation based on the parameter covariance matrix $\gras{C}_M$
\cite{metzger2002,tarantola2005}:
\begin{equation}
\sigma_y^2 = \sum_{ij} G_{yi}(\gras{C}_{M}^{-1})_{ij}G_{yj}
\ \ \ \ 
G_{yi}(\gras{x}) =  \frac{\partial \eta_y}{\partial x_i}(\gras{x}),
\end{equation}
If one now considers two new observables $y$ and $y'$, possibly 
correlated, such as the neutron skin in $^{208}$Pb and electric dipole (E1) 
polarizability $\alpha_{D}$ in the same nucleus, then the above formula should be 
generalized to
\begin{equation}
\gras{C}_{yy'} = \gras{G}^{T}\gras{C}_{M}^{-1}\gras{G}
\end{equation}
to account for cross-correlations.

In the context of DFT applications, such covariance 
analysis has been applied to compare statistical and systematic uncertainties of 
neutron skins \cite{kortelainen2013}; to explore the properties of ground-state 
properties of closed-shell nuclei far from stability \cite{gao2013}; and to optimize 
EDF for nuclear astrophysics \cite{erler2013,paar2014,chen2014}.

\begin{figure}[!ht]
\center
\includegraphics[width=0.95\linewidth]{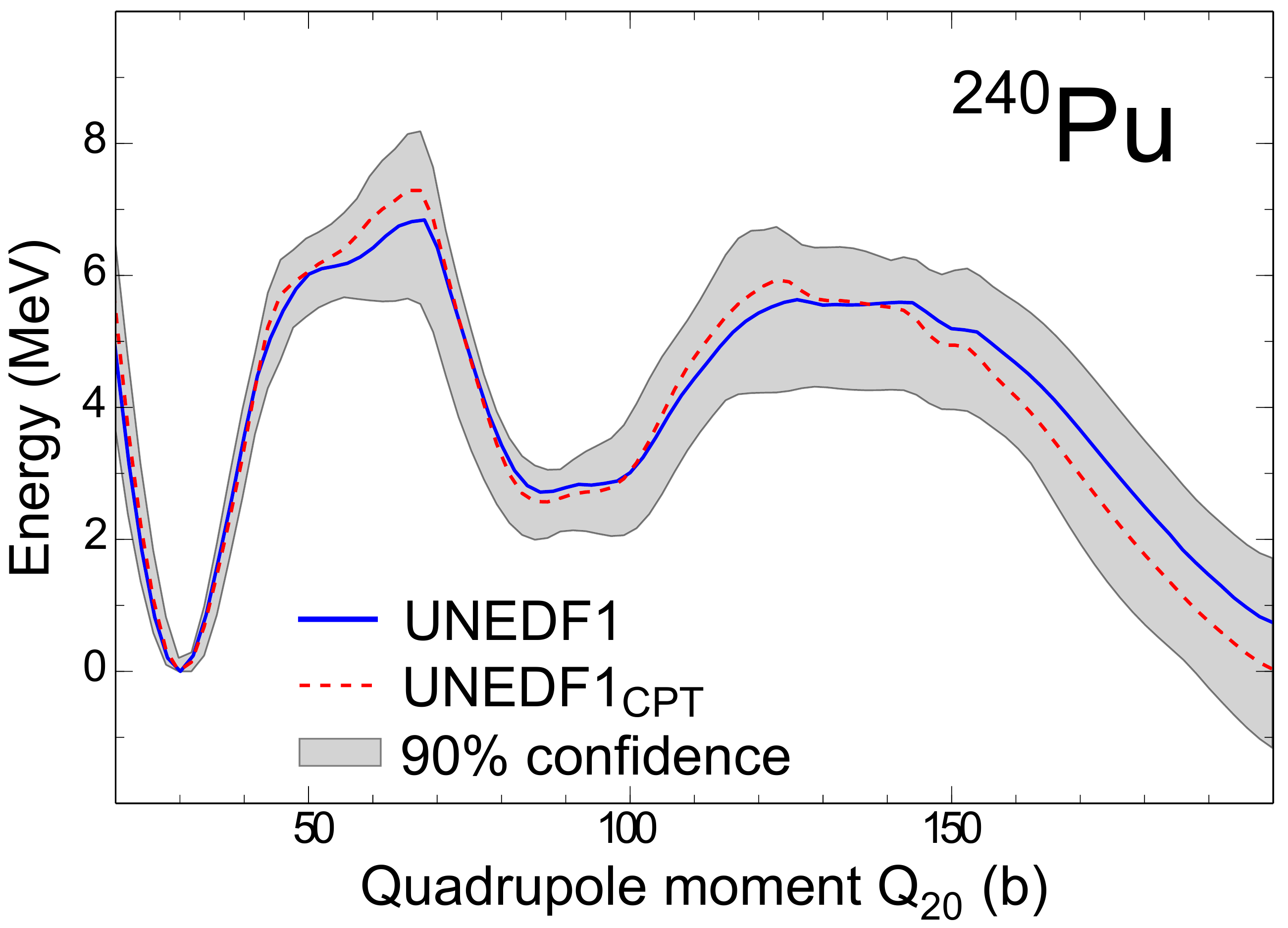}
\caption{(Color online) Comparison between the fission barrier predictions for
$^{240}$Pu made with UNEDF1 (solid line), with a refit of UNEDF1 including 17 more 
masses in neutron-rich nuclei (dashed line), together with the 90\% confidence 
interval (shaded gray area) obtained from Bayesian analysis; from 
\cite{mcdonnell2015}.}
\label{fig:fission}
\end{figure}

Bayesian techniques have been introduced only recently in nuclear theory in 
general, and EDF optimization in particular. As a result, in only a couple 
of cases have these methods been applied to the propagation of uncertainties. In 
\cite{goriely2014}, the backward-forward Monte-Carlo algorithm \cite{bauge2011}, 
which is a particular implementation of Bayesian inference, was used to estimate 
the statistical uncertainties in Skyrme mass models. 
In 
\cite{schunck2015,mcdonnell2015}, the full posterior distribution of the UNEDF1 
Skyrme EDF was determined in a statistical setting by using Bayesian inference, 
with uniform prior for $\gras{x}$ and a Gaussian process to emulate the response 
of the model $\eta(\gras{x})$. The posterior distribution was then sampled and used to estimate 
uncertainties on the fission barrier of $^{240}$Pu and the position of the 
two-neutron dripline. The large uncertainties on fission barriers visible in 
fig.~\ref{fig:fission} emphasizes the lack of constraints on model parameters, 
which could be caused by an inappropriate choice of experimental data 
and/or too limited a model (here Skyrme EDF).

In addition to the applications mentioned in the previous section, a few 
attempts have been made
 to propagate statistical uncertainties from the nuclear EDF 
to the calculation of observables that involved another model. For example, 
quantifying the impact of neutron skins on the electric dipole polarizability 
or on the weak-charge form factor requires calculating  the electric 
dipole response function, that is, RPA calculations \cite{piekarewicz2012,reinhard2013,reinhard2013-a}.

%
%
%
%
\section{Conclusions}
\label{sec:conclusions}

Over the past decade, nuclear density functional theory has positioned itself as 
a candidate for a global, comprehensive, accurate, and predictive theory of 
nuclear structure and reactions. Thanks to the (very recent) introduction in this 
field of standard statistical tools such as covariance techniques or Bayesian 
inference, the statistical uncertainties associated with the most common 
energy functionals such as the Skyrme, Gogny, or relativistic EDF have been 
computed rigorously. The propagation of these uncertainties to model prediction 
in nuclei far from stability has often highlighted the need to substantially 
improve the constraints on the parameters of the nuclear EDF, irrespective of 
the origin of the functional itself. Progress is thus needed in two complementary 
directions. Better rooting of the nuclear EDF in the theory of 
nuclear forces will provide much needed constraints on the expected 
predictive power of the theory. And, this effort should go hand 
in hand with the generalization of statistical techniques to the problem of 
EDF optimization, and the always-indispensable conversations with the experimental 
nuclear physics and data communities.
 
On a practical level, an exciting avenue of research would be to extend the use
of statistical techniques to complex problems where the nuclear EDF is one of 
several theoretical tools used. For example, properties of the neutron spectrum 
in neutron-induced fission are currently described within the Hauser-Feshbach 
approach to nuclear reactions. Such calculations require fission fragment yields, 
total kinetic energies, and excitation energies of the fragments. These 
quantities, in turn, are currently obtained from either semi-phenomenological 
models based, for example, on Langevin dynamics \cite{randrup2011,randrup2011-a}, or 
from fully microscopic time-dependent generator coordinate method calculations 
\cite{goutte2004,younes2012,regnier2015}. Either way, these dynamical 
calculations depend on the potential energy surface of the nucleus in some 
pre-defined collective space. For the microscopic approach, this potential 
energy surface depends on which nuclear EDF is used, how the EDF has 
been fitted, and what types of corrections are included  
\cite{nikolov2011,kortelainen2012,hao2012,mcdonnell2013,giuliani2013,schunck2014}. 
Ultimately, one would therefore wish to propagate the uncertainties all the 
way through this chain of ``models,'' from the nuclear EDF to the fission 
spectrum. 

A related area of future research would be to define a comprehensive 
framework to address uncertainties. In this manuscript, we have insisted on the 
statistical uncertainties, with only a short discussion of numerical errors. 
However, we have also pointed out that all forms of uncertainties are related to 
one another: numerical errors are not a constant offset in DFT calculations, and 
thus they propagate in a very nonlinear way into the calculation of the $\chi^2$, 
which will impact parameter optimization and subsequent uncertainty analysis. The 
particular mathematical formulation of the theory (SR-EDF versus MR-EDF, HFB 
approximation only or HFB plus corrections, etc.) also partially determines 
which observables can be reliably computed by the model. For all others, the 
statistical analysis may reveal that some parameters are ill-constrained, not 
because the data is insufficient, but because the model is not sensitive to it. 
Moreover, one should work toward incorporating experimental uncertainties. In 
the case of the UNEDF2 parameterization, for example, both fission isomer 
excitation energies and single-particle states were included in the fit. Yet 
these quantities are model-dependent, and their ``experimental'' error is rather 
large. In the future, one should try to incorporate this information in the 
determination of EDF parameters.

\begin{acknowledgement}
This material is based upon work supported by the U.S.\ Department of Energy,
Office of Science, Office of Nuclear Physics under award numbers
DE-AC52-07NA27344 (Lawrence Livermore National Laboratory), DE-AC02-06CH11357
(Argonne National Laboratory), and DE-SC0008511 (NUCLEI SciDAC Collaboration),
and by the NNSA's Stewardship Science Academic Alliances Program under award
no.\ DE-NA0001820. Computational resources were provided through an INCITE
award ``Computational Nuclear Structure'' by the National Center for
Computational Sciences and National Institute for Computational Sciences at Oak
Ridge National Laboratory, through an award by the Livermore Computing Resource
Center at Lawrence Livermore National Laboratory, and through an award by the
Laboratory Computing Resource Center at Argonne National Laboratory.
\end{acknowledgement}

\bibliographystyle{unsrt}
\bibliography{zotero_output,books}

\begin{thebibliography}{100}

\bibitem{schmidt2014}
{Karl-Heinz} Schmidt, Beatriz Jurado, and Charlotte Amouroux.
\newblock General description of fission observables.
\newblock Technical Report {NEA/DB/DOC(2014)1}, Organisation for Economic
  {Co-Operation} and Development, Nuclear Energy {Agency-OECD/NEA}, Le Seine
  {Saint-Germain}, 12 boulevard des Iles, F-92130 {Issy-les-Moulineaux}
  {(France)}, 2014.

\bibitem{erler2012}
Jochen Erler, Noah Birge, Markus Kortelainen, Witold Nazarewicz, Erik Olsen,
  Alexander~M. Perhac, and Mario Stoitsov.
\newblock The limits of the nuclear landscape.
\newblock {\em Nature}, 486(7404):509, 2012.

\bibitem{bogner2013}
Scott Bogner, Aurel Bulgac, J.~Carlson, Jonathan Engel, George Fann, Richard~J.
  Furnstahl, Stefano Gandolfi, Gaute Hagen, Mihai Horoi, and C.~Johnson.
\newblock Computational nuclear quantum many-body problem: The {UNEDF} project.
\newblock {\em Comput. Phys. Comm.}, 184(10):2235, 2013.

\bibitem{goriely2013}
S.~Goriely, N.~Chamel, and J.~M. Pearson.
\newblock {Hartree-Fock-Bogoliubov} nuclear mass model with 0.50 {MeV} accuracy
  based on standard forms of {Skyrme} and pairing functionals.
\newblock {\em Phys. Rev. C}, 88(6):061302, 2013.

\bibitem{goriely2009}
S.~Goriely, S.~Hilaire, M.~Girod, and S.~P\'{e}ru.
\newblock First {Gogny-Hartree-Fock-Bogoliubov} nuclear mass model.
\newblock {\em Phys. Rev. Lett.}, 102(24):242501, 2009.

\bibitem{drut2010}
{J.E.} Drut, {R.J.} Furnstahl, and L.~Platter.
\newblock Toward ab initio density functional theory for nuclei.
\newblock {\em Prog. Part. Nucl. Phys.}, 64(1):120, 2010.

\bibitem{stoitsov2010}
M.~Stoitsov, M.~Kortelainen, S.~K. Bogner, T.~Duguet, R.~J. Furnstahl,
  B.~Gebremariam, and N.~Schunck.
\newblock Microscopically based energy density functionals for nuclei using the
  density matrix expansion: Implementation and pre-optimization.
\newblock {\em Phys. Rev. C}, 82(5):054307, 2010.

\bibitem{carlsson2010}
{B.G.} Carlsson, J.~Dobaczewski, J.~Toivanen, and P.~Vesel\'{y}.
\newblock Solution of self-consistent equations for the {N3LO} nuclear energy
  density functional in spherical symmetry. the program hosphe (v1.02).
\newblock {\em Comput. Phys. Comm.}, 181(9):1641, 2010.

\bibitem{duguet2015}
T.~Duguet.
\newblock Symmetry broken and restored coupled-cluster theory: I. rotational
  symmetry and angular momentum.
\newblock {\em J. Phys. G: Nucl. Part. Phys.}, 42(2):025107, 2015.

\bibitem{hohenberg1964}
P.~Hohenberg and W.~Kohn.
\newblock Inhomogeneous electron gas.
\newblock {\em Phys. Rev.}, 136({3B}):B864, 1964.

\bibitem{kohn1965}
W.~Kohn and L.~J. Sham.
\newblock {Self-Consistent} equations including exchange and correlation
  effects.
\newblock {\em Phys. Rev.}, 140({4A}):A1133, 1965.

\bibitem{parr1989}
R.G. Parr and W.~Yang.
\newblock {\em Density Functional Theory of Atoms and Molecules}.
\newblock Oxford University Press, Oxford, 1989.

\bibitem{dreizler1990}
R.M. Dreizler and E.K.U Gross.
\newblock {\em Density Functional Theory: An Approach to the Quantum Many-Body
  Problem}.
\newblock Springer-Verlag, 1990.

\bibitem{eschrig1996}
R.~Eschrig.
\newblock {\em Fundamentals of Density Functional Theory}.
\newblock Teubner, Leipzig, 1996.

\bibitem{bender2003}
Michael Bender, {Paul-Henri} Heenen, and {Paul-Gerhard} Reinhard.
\newblock Self-consistent mean-field models for nuclear structure.
\newblock {\em Rev. Mod. Phys.}, 75(1):121, 2003.

\bibitem{ring2000}
P.~Ring and P.~Schuck.
\newblock {\em The Nuclear Many-Body Problem}.
\newblock Springer-Verlag, 2000.

\bibitem{scheidenberger2014}
T.~Duguet.
\newblock The nuclear energy density functional formalism.
\newblock In Christoph Scheidenberger and Marek Pf\"{u}tzner, editors, {\em The
  Euroschool on Exotic Beams, Vol. {IV}}, volume 879, pages 293--350. Springer
  Berlin Heidelberg, Berlin, Heidelberg, 2014.

\bibitem{blaizot1985}
J.-P. Blaizot and G.~Ripka.
\newblock {\em Quantum Theory of Finite Systems}.
\newblock The MIT Press, Cambridge, 1985.

\bibitem{brink2005}
D.M. Brink and R.A. Broglia, editors.
\newblock {\em Nuclear Superfluidity - Pairing in Finite Systems}.
\newblock Cambridge University Press, 2005.

\bibitem{valatin1961}
J.~G. Valatin.
\newblock Generalized {Hartree-Fock} method.
\newblock {\em Phys. Rev.}, 122(4):1012, 1961.

\bibitem{mang1975}
{Hans-J\"{o}rg} Mang.
\newblock The self-consistent single-particle model in nuclear physics.
\newblock {\em Phys. Rep.}, 18(6):325, 1975.

\bibitem{dobaczewski2000}
J.~Dobaczewski and J.~Dudek.
\newblock Solution of the {Skyrme-Hartree-Fock} equations in the cartesian
  deformed harmonic-oscillator {basis.(III)} {HFODD} (v1. 75r): A new version
  of the program.
\newblock {\em Comput. Phys. Comm.}, 131:164, 2000.

\bibitem{dobaczewski2000-a}
J.~Dobaczewski, J.~Dudek, S.~G. Rohozi\'{n}ski, and T.~R. Werner.
\newblock Point symmetries in the {Hartree-Fock} approach. i. densities,
  shapes, and currents.
\newblock {\em Phys. Rev. C}, 62(1):014310, 2000.

\bibitem{rohoziski2010}
S.~G. Rohozi\'{n}ski, J.~Dobaczewski, and W.~Nazarewicz.
\newblock Self-consistent symmetries in the proton-neutron
  {Hartree-Fock-Bogoliubov} approach.
\newblock {\em Phys. Rev. C}, 81(1):014313, 2010.

\bibitem{skyrme1959}
T.~H.~R. Skyrme.
\newblock The effective nuclear potential.
\newblock {\em Nucl. Phys.}, 9(4):615, 1959.

\bibitem{vautherin1972}
D.~Vautherin and D.~M. Brink.
\newblock {Hartree-Fock} calculations with {Skyrme}'s interaction. i. spherical
  nuclei.
\newblock {\em Phys. Rev. C}, 5(3):626, 1972.

\bibitem{decharge1980}
J.~Decharg\'{e} and D.~Gogny.
\newblock {Hartree-Fock-Bogolyubov} calculations with the d1 effective
  interaction on spherical nuclei.
\newblock {\em Phys. Rev. C}, 21(4):1568, 1980.

\bibitem{stone2007}
{J.R.} Stone and {P.-G.} Reinhard.
\newblock The {Skyrme} interaction in finite nuclei and nuclear matter.
\newblock {\em Prog. Part. Nucl. Phys.}, 58(2):587--657, 2007.

\bibitem{erler2010}
J.~Erler, P.~Kl\"{u}pfel, and {P.-G.} Reinhard.
\newblock Misfits in {Skyrme{\textendash}Hartree{\textendash}Fock}.
\newblock {\em J. Phys. G: Nucl. Part. Phys.}, 37(6):064001, 2010.

\bibitem{kortelainen2010}
M.~Kortelainen, T.~Lesinski, J.~Mor\'{e}, W.~Nazarewicz, J.~Sarich, N.~Schunck,
  M.~V. Stoitsov, and S.~Wild.
\newblock Nuclear energy density optimization.
\newblock {\em Phys. Rev. C}, 82(2):024313, 2010.

\bibitem{kortelainen2012}
M.~Kortelainen, J.~{McDonnell}, W.~Nazarewicz, {P.-G.} Reinhard, J.~Sarich,
  N.~Schunck, M.~V. Stoitsov, and S.~M. Wild.
\newblock Nuclear energy density optimization: Large deformations.
\newblock {\em Phys. Rev. C}, 85(2):024304, 2012.

\bibitem{kortelainen2014}
M.~Kortelainen, J.~{McDonnell}, W.~Nazarewicz, E.~Olsen, {P.-G.} Reinhard,
  J.~Sarich, N.~Schunck, S.~M. Wild, D.~Davesne, J.~Erler, and A.~Pastore.
\newblock Nuclear energy density optimization: Shell structure.
\newblock {\em Phys. Rev. C}, 89(5):054314, 2014.

\bibitem{anguiano2001}
M.~Anguiano, J.~L. Egido, and L.~M. Robledo.
\newblock Coulomb exchange and pairing contributions in nuclear
  {Hartree{\textendash}Fock{\textendash}Bogoliubov} calculations with the
  {Gogny} force.
\newblock {\em Nucl. Phys. A}, 683(1{\textendash}4):227, 2001.

\bibitem{stoitsov2007}
M.~V. Stoitsov, J.~Dobaczewski, R.~Kirchner, W.~Nazarewicz, and J.~Terasaki.
\newblock Variation after particle-number projection for the
  {Hartree-Fock-Bogoliubov} method with the {Skyrme} energy density functional.
\newblock {\em Phys. Rev. C}, 76(1):014308, 2007.

\bibitem{bender2009}
M.~Bender, K.~Bennaceur, T.~Duguet, P.~{-H.} Heenen, T.~Lesinski, and J.~Meyer.
\newblock Tensor part of the {Skyrme} energy density functional. {II.}
  deformation properties of magic and semi-magic nuclei.
\newblock {\em Phys. Rev. C}, 80(6):064302, 2009.

\bibitem{duguet2009}
T.~Duguet, M.~Bender, K.~Bennaceur, D.~Lacroix, and T.~Lesinski.
\newblock Particle-number restoration within the energy density functional
  formalism: Nonviability of terms depending on noninteger powers of the
  density matrices.
\newblock {\em Phys. Rev. C}, 79(4):044320, 2009.

\bibitem{lacroix2009}
D.~Lacroix, T.~Duguet, and M.~Bender.
\newblock Configuration mixing within the energy density functional formalism:
  Removing spurious contributions from nondiagonal energy kernels.
\newblock {\em Phys. Rev. C}, 79(4):044318, 2009.

\bibitem{raimondi2014}
F.~Raimondi, K.~Bennaceur, and J.~Dobaczewski.
\newblock Nonlocal energy density functionals for low-energy nuclear structure.
\newblock {\em J. Phys. G: Nucl. Part. Phys.}, 41(5):055112, 2014.

\bibitem{sadoudi2013}
J.~Sadoudi, T.~Duguet, J.~Meyer, and M.~Bender.
\newblock {Skyrme} functional from a three-body pseudopotential of second order
  in gradients: Formalism for central terms.
\newblock {\em Phys. Rev. C}, 88(6):064326, 2013.

\bibitem{sadoudi2013-a}
J~Sadoudi, M~Bender, K~Bennaceur, D~Davesne, R~Jodon, and T~Duguet.
\newblock {Skyrme} pseudo-potential-based {EDF} parametrization for
  spuriousity-free {MR} {EDF} calculations.
\newblock {\em Phys. Scr.}, T154:014013, 2013.

\bibitem{fayans1994}
S.~A. Fayans, S.~V. Tolokonnikov, E.~L. Trykov, and D.~Zawischa.
\newblock Isotope shifts within the energy-density functional approach with
  density dependent pairing.
\newblock {\em Physics Letters B}, 338(1):1, 1994.

\bibitem{kroemer1995}
E.~Kr\"{o}mer, S.~V. Tolokonnikov, S.~A. Fayans, and D.~Zawischa.
\newblock Energy-density functional approach for non-spherical nuclei.
\newblock {\em Physics Letters B}, 363(1{\textendash}2):12, 1995.

\bibitem{fayans2000}
S.~A. Fayans, S.~V. Tolokonnikov, E.~L. Trykov, and D.~Zawischa.
\newblock Nuclear isotope shifts within the local energy-density functional
  approach.
\newblock {\em Nuclear Physics A}, 676(1{\textendash}4):49, 2000.

\bibitem{baldo2008}
M.~Baldo, P.~Schuck, and X.~Vi\~{n}as.
\newblock {Kohn{\textendash}Sham} density functional inspired approach to
  nuclear binding.
\newblock {\em Physics Letters B}, 663(5):390, 2008.

\bibitem{baldo2013}
M.~Baldo, L.~M. Robledo, P.~Schuck, and X.~Vi\~{n}as.
\newblock New {Kohn-Sham} density functional based on microscopic nuclear and
  neutron matter equations of state.
\newblock {\em Phys. Rev. C}, 87(6):064305, 2013.

\bibitem{hupin2011}
Guillaume Hupin and Denis Lacroix.
\newblock Description of pairing correlation in many-body finite systems with
  density functional theory.
\newblock {\em Phys. Rev. C}, 83(2):024317, 2011.

\bibitem{hupin2011-a}
Guillaume Hupin, Denis Lacroix, and Michael Bender.
\newblock Formulation of functional theory for pairing with particle number
  restoration.
\newblock {\em Phys. Rev. C}, 84(1):014309, 2011.

\bibitem{hupin2012}
Guillaume Hupin and Denis Lacroix.
\newblock Number-conserving approach to the pairing problem: Application to kr
  and sn isotopic chains.
\newblock {\em Phys. Rev. C}, 86(2):024309, 2012.

\bibitem{lesinski2014}
Thomas Lesinski.
\newblock Density functional theory with spatial-symmetry breaking and
  configuration mixing.
\newblock {\em Phys. Rev. C}, 89(4):044305, 2014.

\bibitem{dobaczewski2009}
Jacek Dobaczewski.
\newblock Lipkin translational-symmetry restoration in the mean-field and
  energy{\textendash}density-functional methods.
\newblock {\em J. Phys. G: Nucl. Part. Phys.}, 36(10):105105, 2009.

\bibitem{wang2014}
X.~B. Wang, J.~Dobaczewski, M.~Kortelainen, L.~F. Yu, and M.~V. Stoitsov.
\newblock Lipkin method of particle-number restoration to higher orders.
\newblock {\em Phys. Rev. C}, 90(1):014312, 2014.

\bibitem{engel2007}
J.~Engel.
\newblock Intrinsic-density functionals.
\newblock {\em Phys. Rev. C}, 75(1):014306, 2007.

\bibitem{messud2009}
J\'{e}r\'{e}mie Messud, Michael Bender, and Eric Suraud.
\newblock Density functional theory and {Kohn-Sham} scheme for self-bound
  systems.
\newblock {\em Phys. Rev. C}, 80(5):054314, 2009.

\bibitem{dobaczewski2002}
J.~Dobaczewski, W.~Nazarewicz, and M.~V. Stoitsov.
\newblock Contact pairing interaction for the {Hartree-Fock-Bogoliubov}
  calculations.
\newblock In {\em The Nuclear {Many-Body} Problem 2001}, page 181. Springer,
  2002.

\bibitem{tian2009}
Yuan Tian, Zhong-yu Ma, and Peter Ring.
\newblock Separable pairing force for relativistic quasiparticle random-phase
  approximation.
\newblock {\em Phys. Rev. C}, 79(6):064301, 2009.

\bibitem{tian2009-a}
Yuan Tian, Zhong-yu Ma, and P.~Ring.
\newblock Axially deformed relativistic {Hartree} {Bogoliubov} theory with a
  separable pairing force.
\newblock {\em Phys. Rev. C}, 80(2):024313, 2009.

\bibitem{bogner2011}
S.~K. Bogner, R.~J. Furnstahl, H.~Hergert, M.~Kortelainen, P.~Maris,
  M.~Stoitsov, and J.~P. Vary.
\newblock Testing the density matrix expansion against ab initio calculations
  of trapped neutron drops.
\newblock {\em Phys. Rev. C}, 84(4):044306, 2011.

\bibitem{gandolfi2011}
S.~Gandolfi, J.~Carlson, and Steven~C. Pieper.
\newblock Cold neutrons trapped in external fields.
\newblock {\em Phys. Rev. Lett.}, 106(1):012501, 2011.

\bibitem{metzger2002}
W.J. Metzger.
\newblock Statistical methods in data analysis.
\newblock Technical report, Katholieke Universiteit Nijmegen, 2002.

\bibitem{brandt2014}
Siegmund Brandt.
\newblock {\em Data Analysis - Statistical and Computational Methods for
  Scientists and Engineers}.
\newblock Springer, 2014.

\bibitem{goriely2007}
S.~Goriely, M.~Samyn, and J.~Pearson.
\newblock Further explorations of {Skyrme-Hartree-Fock-Bogoliubov} mass
  formulas. {VII.} simultaneous fits to masses and fission barriers.
\newblock {\em Phys. Rev. C}, 75(6):064312, 2007.

\bibitem{kluepfel2009}
P.~Kl\"{u}pfel, {P.-G.} Reinhard, T.~J. B\"{u}rvenich, and J.~A. Maruhn.
\newblock Variations on a theme by {Skyrme}: A systematic study of adjustments
  of model parameters.
\newblock {\em Phys. Rev. C}, 79(3):034310, 2009.

\bibitem{bartel1982}
J.~Bartel, Ph~Quentin, Matthias Brack, C.~Guet, and {H.-B.} H{\aa}kansson.
\newblock Towards a better parametrisation of {Skyrme}-like effective forces: A
  critical study of the {SkM} force.
\newblock {\em Nucl. Phys. A}, 386(1):79, 1982.

\bibitem{berger1991}
J.~F. Berger, M.~Girod, and D.~Gogny.
\newblock Time-dependent quantum collective dynamics applied to nuclear
  fission.
\newblock {\em Comput. Phys. Comm.}, 63(1):365, 1991.

\bibitem{satula1998}
W.~Satu{\l}a, J.~Dobaczewski, and W.~Nazarewicz.
\newblock Odd-even staggering of nuclear masses: Pairing or shape effect?
\newblock {\em Phys. Rev. Lett.}, 81(17):3599, 1998.

\bibitem{rutz1999}
K.~Rutz, M.~Bender, {P.-G.} Reinhard, and J.~A. Maruhn.
\newblock Pairing gap and polarisation effects.
\newblock {\em Phys. Lett. B}, 468(1):1, 1999.

\bibitem{dobaczewski2001}
J.~Dobaczewski, W.~Nazarewicz, and P.~{-G.} Reinhard.
\newblock Pairing interaction and self-consistent densities in neutron-rich
  nuclei.
\newblock {\em Nucl. Phys. A}, 693(1{\textendash}2):361, 2001.

\bibitem{duguet2001}
T.~Duguet, P.~Bonche, {P.-H.} Heenen, and J.~Meyer.
\newblock Pairing correlations. {II.} microscopic analysis of odd-even mass
  staggering in nuclei.
\newblock {\em Phys. Rev. C}, 65(1):014311, 2001.

\bibitem{bertsch2009}
G.~Bertsch, C.~Bertulani, W.~Nazarewicz, N.~Schunck, and M.~Stoitsov.
\newblock Odd-even mass differences from self-consistent mean field theory.
\newblock {\em Phys. Rev. C}, 79(3):034306, 2009.

\bibitem{tondeur2000}
F.~Tondeur, S.~Goriely, J.~M. Pearson, and M.~Onsi.
\newblock Towards a {Hartree-Fock} mass formula.
\newblock {\em Phys. Rev. C}, 62(2):024308, 2000.

\bibitem{wild2015}
Stefan~M. Wild, Jason Sarich, and Nicolas Schunck.
\newblock Derivative-free optimization for parameter estimation in
  computational nuclear physics.
\newblock {\em J. Phys. G: Nucl. Part. Phys.}, 42(3):034031, 2015.

\bibitem{tao-man}
T.~Munson, J.~Sarich, Stefan~M. Wild, S.~Benson, and L.~{Curfman McInnes}.
\newblock {TAO} 2.0 users manual.
\newblock Tech.\ Memo. ANL/MCS-TM-322, Argonne National Laboratory, Argonne,
  IL, 2012.

\bibitem{SWCHAP14}
Stefan~M. Wild.
\newblock Solving derivative-free nonlinear least squares with {POUNDERS}.
\newblock Preprint ANL/MCS-P5120-0414, Argonne, April 2014.

\bibitem{ekstroem2013}
A.~Ekstr\"{o}m, G.~Baardsen, C.~Forss\'{e}n, G.~Hagen, M.~{Hjorth-Jensen},
  G.~R. Jansen, R.~Machleidt, W.~Nazarewicz, T.~Papenbrock, J.~Sarich, and
  S.~M. Wild.
\newblock Optimized chiral {Nucleon-Nucleon} interaction at
  {Next-to-Next-to-Leading} order.
\newblock {\em Phys. Rev. Lett.}, 110(19):192502, 2013.

\bibitem{bertolli2012}
M.~Bertolli, T.~Papenbrock, and S.~M. Wild.
\newblock Occupation-number-based energy functional for nuclear masses.
\newblock {\em Phys. Rev. C}, 85(1):014322, 2012.

\bibitem{schunck2015}
Nicolas Schunck, Jordan~D. {McDonnell}, Jason Sarich, Stefan~M. Wild, and Dave
  Higdon.
\newblock Error analysis in nuclear density functional theory.
\newblock {\em J. Phys. G: Nucl. Part. Phys.}, 42(3):034024, 2015.

\bibitem{dobaczewski2014}
J.~Dobaczewski, W.~Nazarewicz, and {P.-G.} Reinhard.
\newblock Error estimates of theoretical models: A guide.
\newblock {\em J. Phys. G: Nucl. Part. Phys.}, 41:074001, 2014.

\bibitem{toivanen2008}
J.~Toivanen, J.~Dobaczewski, M.~Kortelainen, and K.~Mizuyama.
\newblock Error analysis of nuclear mass fits.
\newblock {\em Phys. Rev. C}, 78(3):034306, 2008.

\bibitem{kortelainen2008}
M.~Kortelainen, J.~Dobaczewski, K.~Mizuyama, and J.~Toivanen.
\newblock Dependence of single-particle energies on coupling constants of the
  nuclear energy density functional.
\newblock {\em Phys. Rev. C}, 77(6):064307, 2008.

\bibitem{montgomery2002}
D.C. Montgomery and G.C. Runger, editors.
\newblock {\em Applied Statistics and Probability for Engineers}.
\newblock John Wiley \& Sons, Inc., 2002.

\bibitem{dudek2011}
J~Dudek, B~Szpak, {M-G} Porquet, and B~Fornal.
\newblock Statistical significance of theoretical predictions: A new dimension
  in nuclear structure theories {(I)}.
\newblock {\em J. Phys.: Conf. Ser.}, 267:012062, 2011.

\bibitem{szpak2011}
B~Szpak, J~Dudek, {M-G} Porquet, and B~Fornal.
\newblock Statistical significance of theoretical predictions: A new dimension
  in nuclear structure theories {(II)}.
\newblock {\em J. Phys.: Conf. Ser.}, 267:012063, 2011.

\bibitem{froehner2000}
F.~H. Fr\"{o}hner.
\newblock Evaluation and analysis of nuclear resonance data.
\newblock Technical Report~18, {OECD} Nuclear Energy Agency, Paris, 2000.

\bibitem{leeb2008}
H.~Leeb, D.~Neudecker, and Th. Srdinko.
\newblock Consistent procedure for nuclear data evaluation based on modeling.
\newblock {\em Nucl. Data Sheets}, 109(12):2762, 2008.

\bibitem{capote2008}
Roberto Capote and Donald~L. Smith.
\newblock An investigation of the performance of the unified {Monte Carlo}
  method of neutron cross section data evaluation.
\newblock {\em Nucl. Data Sheets}, 109(12):2768, 2008.

\bibitem{herman2011}
M.~Herman and A.~Koning.
\newblock Covariance data in the fast neutron region.
\newblock Technical Report~24, Technical report {NEA/WPEC-24}, {OECD} Nuclear
  Energy Agency, Paris, 2011.

\bibitem{talou2015}
Patrick Talou, Toshihiko Kawano, Mark~B. Chadwick, Denise Neudecker, and
  Michael~E. Rising.
\newblock Uncertainties in nuclear fission data.
\newblock {\em J. Phys. G: Nucl. Part. Phys.}, 42(3):034025, 2015.

\bibitem{schindler2009}
M.~R. Schindler and D.~R. Phillips.
\newblock Bayesian methods for parameter estimation in effective field
  theories.
\newblock {\em Annals of Physics}, 324(3):682--708, 2009.

\bibitem{furnstahl2015}
R.~J. Furnstahl, D.~R. Phillips, and S.~Wesolowski.
\newblock A recipe for {EFT} uncertainty quantification in nuclear physics.
\newblock {\em J. Phys. G: Nucl. Part. Phys.}, 42(3):034028, 2015.

\bibitem{mortensen2005}
J.~J. Mortensen, K.~Kaasbjerg, S.~L. Frederiksen, J.~K. N{\o}rskov, J.~P.
  Sethna, and K.~W. Jacobsen.
\newblock Bayesian error estimation in {Density-Functional} theory.
\newblock {\em Phys. Rev. Lett.}, 95(21):216401, 2005.

\bibitem{tarantola2005}
Albert Tarantola.
\newblock {\em Inverse problem theory and methods for model parameter
  estimation}.
\newblock Society for Industrial and Applied Mathematics, Philadelphia, {PA},
  2005.

\bibitem{gamerman2006}
D.~Gamerman and H.F. Lopes.
\newblock {\em Markov Chain Monte Carlo: Stochastic Simulation for Bayesian
  Inference}.
\newblock Chapman \& Hall/CRC, 2006.

\bibitem{kennedy2001}
Marc~C. Kennedy and Anthony {O'Hagan}.
\newblock Bayesian calibration of computer models.
\newblock {\em Journal of the Royal Statistical Society: Series B
  {(Statistical} Methodology)}, 63(3):425--464, 2001.

\bibitem{higdon2008}
Dave Higdon, James Gattiker, Brian Williams, and Maria Rightley.
\newblock Computer model calibration using {High-Dimensional} output.
\newblock {\em J. Am. Statist. Assoc.}, 103(482):570--583, 2008.

\bibitem{bilionis2013}
Ilias Bilionis, Nicholas Zabaras, Bledar~A. Konomi, and Guang Lin.
\newblock Multi-output separable gaussian process: Towards an efficient, fully
  bayesian paradigm for uncertainty quantification.
\newblock {\em J. Comp. Phys.}, 241:212--239, 2013.

\bibitem{goriely2014}
S.~Goriely and R.~Capote.
\newblock Uncertainties of mass extrapolations in {Hartree-Fock-Bogoliubov}
  mass models.
\newblock {\em Phys. Rev. C}, 89(5):054318, 2014.

\bibitem{mcdonnell2015}
J.D. McDonnell, N.~Schunck, D.~Higdon, J.~Sarich, S.M. Wild, and W.~Nazarewicz.
\newblock Uncertainty quantification for nuclear density functional theory and
  information content of new measurements.
\newblock {\em arXiv:1501.03572}, 2015.

\bibitem{bennaceur2005}
K.~Bennaceur and J.~Dobaczewski.
\newblock Coordinate-space solution of the
  {Skyrme{\textendash}Hartree{\textendash}Fock{\textendash}} bogolyubov
  equations within spherical symmetry. the program {HFBRAD} (v1.00).
\newblock {\em Comput. Phys. Comm.}, 168(2):96, 2005.

\bibitem{dobaczewski1997}
J.~Dobaczewski and J.~Dudek.
\newblock Solution of the {Skyrme-Hartree-Fock} equations in the cartesian
  deformed harmonic oscillator basis i. the method.
\newblock {\em Comput. Phys. Comm.}, 102(1{\textendash}3):166, 1997.

\bibitem{dobaczewski1997-a}
J.~Dobaczewski and J.~Dudek.
\newblock Solution of the {Skyrme-Hartree-Fock} equations in the cartesian
  deformed harmonic oscillator basis {II.} the program {HFODD}.
\newblock {\em Comput. Phys. Comm.}, 102(1{\textendash}3):183, 1997.

\bibitem{dobaczewski2004}
J.~Dobaczewski and P.~Olbratowski.
\newblock Solution of the
  {Skyrme{\textendash}Hartree{\textendash}Fock{\textendash}Bogolyubov}
  equations in the cartesian deformed harmonic-oscillator basis. {(IV)} {HFODD}
  (v2.08i): A new version of the program.
\newblock {\em Comput. Phys. Comm.}, 158(3):158, 2004.

\bibitem{dobaczewski2005}
J.~Dobaczewski and P.~Olbratowski.
\newblock Solution of the
  {Skyrme{\textendash}Hartree{\textendash}Fock{\textendash}Bogolyubov}
  equations in the cartesian deformed harmonic-oscillator basis. {(V)}
  {HFODD(v2.08k)}.
\newblock {\em Comput. Phys. Comm.}, 167(3):214, 2005.

\bibitem{stoitsov2005}
{M.V.} Stoitsov, J.~Dobaczewski, W.~Nazarewicz, and P.~Ring.
\newblock Axially deformed solution of the
  {Skyrme{\textendash}Hartree{\textendash}Fock{\textendash}Bogolyubov}
  equations using the transformed harmonic oscillator basis. the program
  {HFBTHO} (v1.66p).
\newblock {\em Comput. Phys. Comm.}, 167(1):43, 2005.

\bibitem{schunck2012}
N.~Schunck, J.~Dobaczewski, J.~{McDonnell}, W.~Satu{\l}a, {J.A.} Sheikh,
  A.~Staszczak, M.~Stoitsov, and P.~Toivanen.
\newblock Solution of the
  {Skyrme{\textendash}Hartree{\textendash}Fock{\textendash}Bogolyubov}
  equations in the cartesian deformed harmonic-oscillator basis. {(VII)}
  {HFODD} (v2.49t): A new version of the program.
\newblock {\em Comput. Phys. Comm.}, 183(1):166, 2012.

\bibitem{stoitsov2013}
{M.V.} Stoitsov, N.~Schunck, M.~Kortelainen, N.~Michel, H.~Nam, E.~Olsen,
  J.~Sarich, and S.~Wild.
\newblock Axially deformed solution of the
  {Skyrme-Hartree{\textendash}Fock{\textendash}Bogoliubov} equations using the
  transformed harmonic oscillator basis {(II)} hfbtho v2.00d: A new version of
  the program.
\newblock {\em Comput. Phys. Comm.}, 184(6):1592, 2013.

\bibitem{schunck2013}
Nicolas Schunck.
\newblock Microscopic description of induced fission.
\newblock {\em J. Phys.: Conf. Ser.}, 436(1):012058, 2013.

\bibitem{dobaczewski1984}
J.~Dobaczewski, H.~Flocard, and J.~Treiner.
\newblock {Hartree-Fock-Bogolyubov} description of nuclei near the neutron-drip
  line.
\newblock {\em Nucl. Phys. A}, 422(1):103, 1984.

\bibitem{pei2008}
J.~Pei, M.~Stoitsov, G.~Fann, W.~Nazarewicz, N.~Schunck, and F.~Xu.
\newblock Deformed coordinate-space {Hartree-Fock-Bogoliubov} approach to
  weakly bound nuclei and large deformations.
\newblock {\em Phys. Rev. C}, 78(6):064306, 2008.

\bibitem{bonche2005}
P.~Bonche, H.~Flocard, and P.~H. Heenen.
\newblock Solution of the {Skyrme} {HF}~+~{BCS} equation on a {3D} mesh.
\newblock {\em Comput. Phys. Comm.}, 171(1):49, 2005.

\bibitem{ryssens2015}
W.~Ryssens, V.~Hellemans, M.~Bender, and P.~{-H.} Heenen.
\newblock Solution of the {Skyrme{\textendash}HF+BCS} equation on a {3D} mesh,
  {II:} a new version of the ev8 code.
\newblock {\em Computer Physics Communications}, 187:175, 2015.

\bibitem{poschl1997}
W~P{\"o}schl, D~Vretenar, A~Rummel, and P~Ring.
\newblock Application of finite element methods in relativistic mean-field
  theory: Spherical nucleus.
\newblock {\em Comput. Phys. Comm.}, 101(1{\textendash}2):75, 1997.

\bibitem{poschl1997-a}
W.~P{\"o}schl, D.~Vretenar, and P.~Ring.
\newblock Relativistic {Hartree-Bogoliubov} theory in coordinate space: Finite
  element solution for nuclear system with spherical symmetry.
\newblock {\em Comput. Phys. Comm.}, 103:217, 1997.

\bibitem{pei2014}
J.~C. Pei, G.~I. Fann, R.~J. Harrison, W.~Nazarewicz, Yue Shi, and S.~Thornton.
\newblock Adaptive multi-resolution {3D} {Hartree-Fock-Bogoliubov} solver for
  nuclear structure.
\newblock {\em Phys. Rev. C}, 90(2):024317, 2014.

\bibitem{samyn2005}
M.~Samyn, S.~Goriely, and J.~Pearson.
\newblock Further explorations of {Skyrme-Hartree-Fock-Bogoliubov} mass
  formulas. v. extension to fission barriers.
\newblock {\em Phys. Rev. C}, 72(4):044316, 2005.

\bibitem{hilaire2007}
S.~Hilaire and M.~Girod.
\newblock Large-scale mean-field calculations from proton to neutron drip lines
  using the {D1S} {Gogny} force.
\newblock {\em Eur. Phys. J. A}, 33(2):237, 2007.

\bibitem{coon2012}
S.~A. Coon, M.~I. Avetian, M.~K.~G. Kruse, U.~van Kolck, P.~Maris, and J.~P.
  Vary.
\newblock Convergence properties of ab initio calculations of light nuclei in a
  harmonic oscillator basis.
\newblock {\em Phys. Rev. C}, 86(5):054002, 2012.

\bibitem{furnstahl2012}
R.~J. Furnstahl, G.~Hagen, and T.~Papenbrock.
\newblock Corrections to nuclear energies and radii in finite oscillator
  spaces.
\newblock {\em Phys. Rev. C}, 86(3):031301, 2012.

\bibitem{more2013}
S.~N. More, A.~Ekstr\"{o}m, R.~J. Furnstahl, G.~Hagen, and T.~Papenbrock.
\newblock Universal properties of infrared oscillator basis extrapolations.
\newblock {\em Phys. Rev. C}, 87(4):044326, 2013.

\bibitem{kortelainen2013}
M.~Kortelainen, J.~Erler, W.~Nazarewicz, N.~Birge, Y.~Gao, and E.~Olsen.
\newblock Neutron-skin uncertainties of {Skyrme} energy density functionals.
\newblock {\em Phys. Rev. C}, 88(3):031305, 2013.

\bibitem{gao2013}
Y.~Gao, J.~Dobaczewski, M.~Kortelainen, J.~Toivanen, and D.~Tarpanov.
\newblock Propagation of uncertainties in the {Skyrme}
  energy-density-functional model.
\newblock {\em Phys. Rev. C}, 87(3):034324, 2013.

\bibitem{erler2013}
J.~Erler, C.~J. Horowitz, W.~Nazarewicz, M.~Rafalski, and {P.-G.} Reinhard.
\newblock Energy density functional for nuclei and neutron stars.
\newblock {\em Phys. Rev. C}, 87(4):044320, 2013.

\bibitem{paar2014}
N.~Paar, Ch.~C. Moustakidis, T.~Marketin, D.~Vretenar, and G.~A. Lalazissis.
\newblock Neutron star structure and collective excitations of finite nuclei.
\newblock {\em Phys. Rev. C}, 90(1):011304, 2014.

\bibitem{chen2014}
{Wei-Chia} Chen and J.~Piekarewicz.
\newblock Building relativistic mean field models for finite nuclei and neutron
  stars.
\newblock {\em Phys. Rev. C}, 90(4):044305, 2014.

\bibitem{bauge2011}
E.~Bauge and P.~{Dossantos-Uzarralde}.
\newblock Evaluation of the covariance matrix of {239Pu} neutronic cross
  sections in the continuum using the {Backward-Forward} {Monte-Carlo} method.
\newblock {\em J. Kor. Phys. Soc.}, 59(23):1218, 2011.

\bibitem{piekarewicz2012}
J.~Piekarewicz, B.~K. Agrawal, G.~Col\`{o}, W.~Nazarewicz, N.~Paar, {P.-G.}
  Reinhard, X.~{Roca-Maza}, and D.~Vretenar.
\newblock Electric dipole polarizability and the neutron skin.
\newblock {\em Phys. Rev. C}, 85(4):041302, 2012.

\bibitem{reinhard2013}
{P.-G.} Reinhard, J.~Piekarewicz, W.~Nazarewicz, B.~K. Agrawal, N.~Paar, and
  X.~{Roca-Maza}.
\newblock Information content of the weak-charge form factor.
\newblock {\em Phys. Rev. C}, 88(3):034325, 2013.

\bibitem{reinhard2013-a}
{P.-G.} Reinhard and W.~Nazarewicz.
\newblock Information content of the low-energy electric dipole strength:
  Correlation analysis.
\newblock {\em Phys. Rev. C}, 87(1):014324, 2013.

\bibitem{randrup2011}
J{\o}rgen Randrup and Peter M\"{o}ller.
\newblock Brownian shape motion on five-dimensional potential-energy surfaces:
  Nuclear fission-fragment mass distributions.
\newblock {\em Phys. Rev. Lett.}, 106(13):132503, 2011.

\bibitem{randrup2011-a}
J.~Randrup, P.~M\"{o}ller, and A.~J. Sierk.
\newblock Fission-fragment mass distributions from strongly damped shape
  evolution.
\newblock {\em Phys. Rev. C}, 84(3):034613, 2011.

\bibitem{goutte2004}
H.~Goutte, P.~Casoli, and J.~{-F.} Berger.
\newblock Mass and kinetic energy distributions of fission fragments using the
  time dependent generator coordinate method.
\newblock {\em Nucl. Phys. A}, 734:217, 2004.

\bibitem{younes2012}
W.~Younes and D.~Gogny.
\newblock Collective dissipation from saddle to scission in a microscopic
  approach.
\newblock Technical Report {LLNL-TR-586694}, Lawrence Livermore National
  Laboratory {(LLNL)}, Livermore, {CA}, 2012.

\bibitem{regnier2015}
D.~Regnier, M.~Verri{\`e}re, N.~Dubray, and N.~Schunck.
\newblock Felix-1.0: A finite element solver for the time-dependent {GCM+GOA}
  equation.
\newblock {\em In preparation}, 2015.

\bibitem{nikolov2011}
N.~Nikolov, N.~Schunck, W.~Nazarewicz, M.~Bender, and J.~Pei.
\newblock Surface symmetry energy of nuclear energy density functionals.
\newblock {\em Phys. Rev. C}, 83(3):034305, 2011.

\bibitem{hao2012}
T.~V.~Nhan Hao, P.~Quentin, and L.~Bonneau.
\newblock Parity restoration in the highly truncated diagonalization approach:
  Application to the outer fission barrier of {240Pu}.
\newblock {\em Phys. Rev. C}, 86(6):064307, 2012.

\bibitem{mcdonnell2013}
J.~D. {McDonnell}, W.~Nazarewicz, and J.~A. Sheikh.
\newblock Third minima in thorium and uranium isotopes in a self-consistent
  theory.
\newblock {\em Phys. Rev. C}, 87(5):054327, 2013.

\bibitem{giuliani2013}
Samuel~A. Giuliani and Luis~M. Robledo.
\newblock Fission properties of the {BCPM} functional.
\newblock {\em Phys. Rev. C}, 88:054325, 2013.

\bibitem{schunck2014}
N.~Schunck, D.~Duke, H.~Carr, and A.~Knoll.
\newblock Description of induced nuclear fission with {Skyrme} energy
  functionals: Static potential energy surfaces and fission fragment
  properties.
\newblock {\em Phys. Rev. C}, 90(5):054305, 2014.

\end{thebibliography}

\end{document}